\documentclass[11pt]{article}
\usepackage{amsfonts}
\usepackage{amsfonts}
\usepackage{amsfonts,amsmath,amssymb,epsf}
\usepackage{graphicx,color}
\usepackage[usenames,dvipsnames]{pstricks}
\usepackage{epsfig}
\usepackage{pst-grad}
\usepackage{pst-plot}
\usepackage{verbatim}
\usepackage{subfigure}
\usepackage{ifthen}
\usepackage{tocvsec2}

\numberwithin{equation}{section}

\newcommand{\al}{\alpha'}
\newcommand{\de}{\partial}
\newcommand{\be}{\begin{equation}}
\newcommand{\ba}{\begin{eqnarray}}
\newcommand{\ea}{\end{eqnarray}}
\newcommand{\ee}{\end{equation}}

\newcommand{\we}{\wedge}

\newcommand{\f}{\frac}
\newcommand{\s}{\sqrt}
\newcommand{\vp}{\varphi}

\newcommand{\ti}{\tilde}
\newcommand{\ap}{\alpha}

\newcommand{\ddd}{\cdot\cdot\cdot}
\newcommand{\no}{\nonumber \\}
\newcommand{\la}{\langle}
\newcommand{\lb}{\rangle}
\newcommand{\ep}{\epsilon}

\newcommand{\x}{\times}
\newcommand{\non}{\nonumber}

\begin{document}
\maxtocdepth{subsection}

\begin{titlepage}
\thispagestyle{empty}

\begin{flushright}
KUNS-2205\\
IPMU09-0055
\end{flushright}

\bigskip

\begin{center}
\noindent{\Large \textbf{On String Theory Duals of Lifshitz-like Fixed Points}}\\
\vspace{15mm} \
Tatsuo Azeyanagi$^{a}$\footnote{e-mail: aze@gauge.scphys.kyoto-u.ac.jp},
 Wei Li$^{b}$\footnote{e-mail:
wei.li@ipmu.jp} and
Tadashi Takayanagi$^{b}$\footnote{e-mail: tadashi.takayanagi@ipmu.jp}\\
\vspace{1cm}

{\it $^{a}$Department of Physics, Kyoto University, Kyoto 606-8502, Japan \\
 $^{b}$Institute for the Physics and Mathematics of the Universe (IPMU), \\
 University of Tokyo, Kashiwa, Chiba 277-8582, Japan}

\vskip 3em
\end{center}

\begin{abstract}
We present type IIB supergravity solutions which are expected to be
dual to certain Lifshitz-like fixed points with anisotropic scale
invariance. They are expected to describe a class of D3-D7 systems
and their finite temperature generalizations are straightforward. We
show that there exist solutions that interpolate between these
anisotropic solutions in the IR and the standard $AdS_5$ solutions
in the UV. This predicts anisotropic RG flows from familiar
isotropic fixed points to anisotropic ones. In our case, these RG
flows are triggered by a non-zero theta-angle in Yang-Mills theories
that linearly depends on one of the spatial coordinates. We study
the perturbations around these backgrounds and discuss the
possibility of instability. We also holographically compute their
thermal entropies, viscosities, and entanglement entropies.

\end{abstract}

\end{titlepage}

\newpage

\tableofcontents

\newpage

\section{Introduction}
\settocdepth{subsection}

The AdS/CFT correspondence has provided a very powerful and
successful paradigm to analyze relativistic and isotropic fixed
points in various quantum field theories
\cite{Maldacena,ADSGKP,ADSWitten,adsreview}. On the field theory
side, they are described by $(d+1)$-dimensional conformal field
theories and are invariant under the homogeneous scaling
transformation $(t,x_1,x_2,\dots,x_d) \to  (\lambda t,\lambda x_1,
\lambda x_2,\dots,\lambda x_d)$. On the gravity side, they are
equivalently described by gravity on a $(d+2)$-dimensional AdS space
\be ds^2=r^2\left(-dt^2+\sum_{i=1}^d dx^2_i\right)+\f{dr^2}{r^2}.
\label{AdSm} \ee

It is natural to try to extend the AdS/CFT correspondence 
to a holography for the
following anisotropic spacetime
\begin{equation}
ds^2=r^{2z}\left(-dt^2+\sum_{i=1}^p dx^2_i\right)+r^2\sum_{j=p+1}^d
dy^2_j +\f{dr^2}{r^2}, \label{AdSani}
\end{equation}
where $0\leq p \leq d-1$, and the parameter $z$($\neq 1$) measures
the degree of Lorentz symmetry violation and anisotropy. Since the
metric (\ref{AdSani}) is invariant under the scaling
$(t,x_i,y_j,r)\to (\lambda^z t,\lambda^z x_i, \lambda y_j,
\frac{r}{\lambda})$, we expect that on the field theory side it is
dual to a fixed point which is invariant under the scaling
transformation \be (t,x_i,y_j)\to (\lambda^z t, \lambda^z x_i,
\lambda y_j).\label{aniscale} \ee Notice also that by a coordinate
redefinition $r^z=\rho$, we can rewrite the metric (\ref{AdSani})
into another illuminating form (after rescaling $(t,x_i,y_i)$
accordingly) \be ds^2=\rho^2\left(-dt^2+\sum_{i=1}^p
dx^2_i\right)+\rho^{\f{2}{z}}\sum_{j=p+1}^d dy^2_j
+\f{d\rho^2}{\rho^2}.\label{aniso} \ee Thus we can equally argue
that the dual background is invariant under an anisotropic scaling
transformation $(t,x_i,y_j,\rho)\to (\lambda t,\lambda x_i,
\lambda^{\frac{1}{z}} y_j, \frac{\rho}{\lambda})$, where the $y_j$
directions are responsible for the Lorentz symmetry violation and anisotropy.

In general, fixed points with the anisotropic scaling property
(\ref{aniscale}) are called Lifshitz(-like) fixed points\footnote{
The most standard example is the free scalar field theory with
$z=2$, known as the Lifshitz model. Anisotropic fixed points in
interacting field theories in general can have $z\neq 2$
\cite{Diehl,Henkel}. Even though the original Lifshitz fixed points
were found in anisotropic magnets where three critical lines meet,
in this paper we simply define Lifshitz-like fixed points as any
fixed points which have anisotropic scale invariance.}
  (see e.g. the
textbook \cite{Cardy} for a brief review). This generalization of
AdS/CFT correspondence to Lifshitz-like fixed points (\ref{AdSani})
was first proposed and analyzed by Kachru, Liu and Mulligan
\cite{Ka} in the particular case of $p=0$. The simplest case with
$p=0$ represents non-relativistic fixed points with dynamical
critical exponent $z$, which appear in many examples of quantum
criticality in condensed matter physics (see the references in
\cite{Ka}). See
\cite{Pal,Horava,Taylor:2008tg,Danielsson:2009gi,PalA,Gordeli:2009vh} for
further progress on holographic aspects of this topic.\footnote{
Gravity duals of another types of fixed points with non-relativistic
scaling symmetry \cite{NRA} have also been studied especially for
systems with non-relativistic conformal invariance
\cite{NRB,NRC,NRD,NRE,NRF,HaYo}.}

The other cases where $1\leq p\leq d-1$ are not only generalizations
of $p=0$ case but can also be interpreted as space-like anisotropic
fixed points (see also \cite{PalA}) as is clear from the expression
(\ref{aniso}). Lifshitz fixed points with space-like anisotropic
scale invariance
appear in realistic magnets such as MnP and the
axial next-nearest-neighbor Ising model \cite{Diehl}. They are also
realized in models of directed percolation \cite{Cardy}.

To understand holographic duals of such gravity backgrounds, it is
the best to embed them into string theory, where microscopic
interpretations are often possible by using D-branes. However, so
far there has been no known embedding of (\ref{AdSani}) in string
theory. Motivated by this circumstance, in this paper we will
construct such anisotropically scaling solutions in type IIB
supergravity. We mainly focus on the backgrounds generated by
intersections of D3 and D7 branes. They correspond to the choice
$p=2$ and $d=3$ and are expected to be non-supersymmetric. This
restriction is imposed not only for the tractability of the
supergravity analysis, but is also due to another motivation, namely
to construct back-reacted D3-D7 solutions that are dual to the pure
Chern-Simons gauge theory in the second setup of \cite{FLRT}. In the
end, we find a class of solutions with the exponent $z=3/2$. We also
extend them to black brane solutions dual to finite temperature
theories.

Furthermore, we show that there exist solutions which interpolate
between our anisotropic solutions and the familiar $AdS_5\times X_5$
solutions. We also construct their numerical solutions. The holography
suggests that our Lifshitz-like fixed points can be obtained from
various four-dimensional CFTs including ${\cal N}=4$ super
Yang-Mills via RG flows\footnote{Here the Lifshitz-like fixed
points are realized in the IR limit. It is also intriguing to
consider opposite RG flows, where IR fixed points become
relativistic and isotropic $z=1$ as in \cite{Ka,Horava}.}. These flows are triggered by the relevant
and anisotropic perturbation which gives a non-zero $\theta$ parameter
(i.e. the coefficient in front of the topological Yang-Mills coupling
$F\we F$) that depends linearly on one of the three spatial
coordinates i.e. $\theta\propto x_3$. Notice that when $x_3$ is
compactified, the perturbation induces the Chern-Simons coupling
$\int A\we F+\f{2}{3}A^3$ as in \cite{FLRT}, which becomes relevant in the IR.

This paper is organized as follows. In section \ref{sec:dual}, we present
solutions dual to a class of Lifshitz-like fixed points based on
D3-D7 systems with their black brane
generalizations. In section \ref{sec:RG}, we show there exist interpolating
solutions which approach the Lifshitz-like scaling solutions in the
IR and the standard $AdS_5$ solutions in the UV.
In section \ref{sec:viscosity}, we
holographically calculate the shear and bulk viscosity.
In section \ref{sec:EE}, we compute their holographic entanglement entropies
and discuss how the scaling behaviors of the entanglement entropies
depend on the direction along which the sub-systems are delineated.
In section \ref{sec:Pert}, we study the perturbations around these backgrounds
and discuss the instabilities. In section \ref{sec:dfds}, we
present anisotropic solutions based on D4-D6 systems. In section
\ref{sec:conc}, we summarize our conclusions.

\section{Holographic Duals of Lifshitz-like Fixed Points in Type IIB String}
\label{sec:dual}

In this section we will present the main result of this paper. We
will construct new solutions in type IIB supergravity with RR 5-form
and 1-form fluxes whose Einstein metrics enjoy a nice scaling
property. Since their scaling is anisotropic as opposed to the
well-known $AdS_5$ background, we argue that they are dual to
Lifshitz-like fixed points described by certain D3-D7 systems.

\subsection{Type IIB Supergravity}

The IIB supergravity action $S_{IIB}=\f{1}{2\kappa^2_{10}}\int {\cal
L}$ in the string frame is defined by the Lagrangian (we follow the
convention in \cite{Pol})
\begin{eqnarray} && {\cal
L}=\s{-g}e^{-2\phi}(R+4\de_M\phi\de^M\phi)-\f{e^{-2\phi}}{2}H_3\we
*H_3 -\f{1}{2}F_1\we *F_1-\f{1}{2}\ti{F}_3\we *\ti{F}_3\no
&&-\f{1}{4}\ti{F}_5\we *\ti{F}_5-\f{1}{2}C_4\we H_3\we F_3,
\label{sugra_action}
\end{eqnarray}
where $F_1=d\chi$, $\ti{F_3}\equiv F_3-\chi H_3$, and
$\ti{F}_5\equiv F_5-\f{1}{2}C_2\we H_3+\f{1}{2}B_2\we F_3$. We set
$\al=1$ therefore $2\kappa^2_{10}=(2\pi)^7$.

The fluxes obey the equations of motion:
\begin{eqnarray}
&& d*F_1=*\ti{F}_3\we H_3,\qquad  d*\ti{F}_3=-H_3\we \ti{F}_5,\qquad
d*\ti{F}_5=H_3\we \ti{F}_3,\no
&& d(e^{-2\phi}*H_3)=F_1\we
*\ti{F}_3+\ti{F}_3\we \ti{F}_5. \label{IIBflux} \end{eqnarray}
plus the Bianchi identities:
\begin{eqnarray} &&
dH_3=0,\qquad dF_1=0,\qquad d\ti{F}_3=H_3\we F_1,\qquad
d\ti{F}_5=H_3\we \ti{F}_3,
\label{IIBfluxBI} \end{eqnarray} and the
self-dual constraint for $\ti{F}_5$:
\begin{equation}
*\ti{F}_5=\ti{F}_5.
\end{equation}

The dilaton equation of motion is \be R+4\nabla_M\nabla^M
\phi-\f{1}{12}H_{M N P}H^{MNP}-4\nabla_M\phi\nabla^M\phi=0.
\label{IIBdilaton} \ee And the Einstein equation becomes \ba &&
R_{MN}+2\nabla_M\nabla_N\phi+\f{1}{4}g_{MN}A\no && \ \ \
=\f{1}{4}H_{MAB}H_{N}^{\ \ AB}+\f{1}{2}e^{2\phi}F_{M}F_N+\f{1}{4}e^{2\phi}
\ti{F}_{MAB}\ti{F}_N^{\ \ AB}+\f{1}{4\cdot 4!}e^{2\phi}
\ti{F}_{MABCD}\ti{F}_N^{\ \ ABCD}, \nonumber\\ \label{IIBEinstein} \ea where \be
A\equiv e^{2\phi}\de_M\chi\de^M\chi+\f{1}{3!}e^{2\phi}
\ti{F}_{ABC}\ti{F}^{ABC}+\f{1}{2\cdot 5!}e^{2\phi}
\ti{F}_{ABCDE}\ti{F}^{ABCDE}. \ee

\subsection{D3-D7 Ansatz}

We start with the (string frame) metric ansatz that preserves the
three-dimensional Lorentz symmetry $SO(2,1)$: \be
ds^2=e^{2b(r)}(-dt^2+dx^2+dy^2)+e^{2h(r)+2a(r)}dw^2+e^{2c(r)-2a(r)}
dr^2+e^{2c(r)}r^2ds^2_{X_5}.\label{ansatzmetric} \ee We require the
five-dimensional compact manifold $X_5$ to be a unit-radius Einstein
manifold with the same Ricci curvature as the unit-radius $S^5$,
i.e. it satisfies \be R_{\ap\beta}=4g_{\ap\beta}. \ee The simplest
example of $X_5$ is obviously the unit radius sphere $S^5$. The
self-dual 5-form and 1-from fluxes are given in terms of constants
$\ap$ and $\beta$ by \ba
&& F_5=\ap\left( \Omega_{X_5}+*\Omega_{X_5}\right), \label{fivef} \\
&& F_1=d\chi=\beta dw, \label{onef} \ea where $\chi$ is the axion
field (i.e. the RR 0-form potential) and $\Omega_{X_5}$ is the
volume form of $X_5$. The fluxes (\ref{fivef}) and (\ref{onef})
satisfy the equations of motion (\ref{IIBflux}). We also assume that
the dilaton $\phi$ only depends on $r$ and both 3-form fluxes ($H_3$
and $F_3$) vanish. Our ansatz, which looks rather different from
 \cite{Ka}, is motivated in part by an attempt to
construct back-reacted solutions of the D3-D7 intersecting systems introduced in
\cite{FLRT}, as will be explained in detail later.

Under this ansatz, the equations of motion for the metric and the
dilaton ((\ref{IIBEinstein}) and (\ref{IIBdilaton})) are summarized
as follows: \ba &&[b'e^{2z}]'=\f{\beta^2}{4}e^{-2a-h+3b+6c}r^{5}
+\f{\ap^2}{4}e^{-4c+3b+h}r^{-5},\no
&&[(a+h)'e^{2z}]'=-\f{\beta^2}{4}e^{-2a-h+3b+6c}r^{5}
+\f{\ap^2}{4}e^{-4c+3b+h}r^{-5},\no &&[(c+\log
r)'e^{2z}]'=\f{4}{r^2}e^{2z-2a}
+\f{\beta^2}{4}e^{-2a-h+3b+6c}r^{5}-\f{\ap^2}{4}e^{-4c+3b+h}r^{-5},\no
&&[(2z+c-a)'e^{2z}]'=\f{20}{r^2}e^{2z-2a}-\f{\beta^2}{4}e^{-2a-h+3b+6c}r^{5}
-\f{\ap^2}{4}e^{-4c+3b+h}r^{-5},\no &&
2z''+c''-a''+2(z')^2+\f{1}{2}(h')^2+a'h'+2(c')^2+(\f{5}{r}+a')c'
+\f{3}{2}(b')^2-\f{10e^{-2a}}{r^2}+\f{5}{2r^2}=0.\nonumber \ea Here
we have defined \be z\equiv\f32b+\f52\log r +a+2c+\f12h-\phi. \ee
The derivative of a function $f$ with respect to $r$ is denoted by
$f'(r)$. An
observation, which will be useful in the next section, is that a
linear combination of the first four equations gives
\begin{equation}
[(2b-2a-\phi-2h)'e^{2z}]'=0.\label{constra}
\end{equation}

\subsection{D3-D7 Scaling Solutions: Holographic Duals
of Lifshitz-like Fixed Points}

Since we are looking for scaling solutions (namely solutions
invariant under scale transformations), we require all metric
components in (\ref{ansatzmetric}) to be power functions of $r$. In
other words, the functions $a,b,c,z$ and $\phi$ are all logarithmic
functions of $r$. For such a scaling ansatz, the equations of motion
$(\ref{IIBdilaton})$ and $(\ref{IIBEinstein})$ reduce to algebraic
equations and the solution is easily found to be:
\begin{eqnarray}
&& a(r)=\f{1}{2}\log\f{12}{11}-\log\xi_s,\ \ \ b(r)=\f{7\xi_s}{6}\log
r+b_0, \ \ \ c(r)=\left(-1+\f{\xi_s}{6}\right)\log r+c_0, \no &&
h(r)=\f{5\xi_s}{6}\log r+\log\xi_s +h_0,\ \ \ \phi(r)=\f{2\xi_s}{3}\log
r +\phi_0, \no && \ap=4 e^{4c_0-\phi_0}, \ \ \ \
\beta=4\s{\f{2}{11}}e^{h_0-c_0-\phi_0},
\end{eqnarray}
where $b_0,c_0,h_0,\phi_0$ and $\xi_s$ are arbitrary constants.
$\xi_s$ corresponds to the degrees of freedom of the
reparameterization of $r$, while $b_0$ and $h_0$ correspond to the
rescaling of the $(t,x,y,w)$ directions.

Without loss of generality, we choose
\begin{equation}
\xi_s=1,\qquad b_0=c_0+\f{1}{2}\log\f{11}{12}, \qquad
h_0=c_0+\log\f{11}{12},
\end{equation}
and the solution in the string frame reads:
\begin{equation}
ds^2_s=\ti{R}_s^2\left[r^{\frac{7}{3}}(-dt^2+dx^2
+dy^2)+r^{\f{5}{3}}dw^2+\f{dr^2}{r^{\frac{5}{3}}}\right]
+R_s^2r^{\frac{1}{3}}ds_{X_5}^2, \label{scalingmetricSF}
\end{equation}
where $R_s^2=\f{12}{11}\ti{R}_s^2= e^{2c_0}$. And the dilaton scales
with $r$ as
\begin{equation}
e^{\phi}=r^{\f{2}{3}}e^{\phi_0}, \label{pdilaton}
\end{equation}
where $e^{\phi_0}=\frac{\sqrt{22}}{3\beta}$.

Since the dilaton depends on $r$ non-trivially, it is helpful to
discuss the metric in the Einstein frame. Indeed, later we will see
explicitly that a large class of scalar fluctuations around this
solution can be described by Klein-Gordon equations on curved
spacetimes based on the Einstein frame metric instead of on the
string frame metric. The above solution in the Einstein frame is
\begin{equation}
ds^2_E=\ti{R}^2\left[r^2(-dt^2+dx^2
+dy^2)+r^{\f{4}{3}}dw^2+\f{dr^2}{r^2}\right] +R^2ds_{X_5}^2,
\label{scalingmetric}
\end{equation} where the radii
\begin{equation}
R^2=\f{12}{11}\ti{R}^2=
e^{-\f{\phi_0}{2}+2c_0}=\frac{\sqrt{\alpha}}{2}.
\end{equation}

The metric (\ref{scalingmetric}) is invariant under the scaling
\begin{equation}
(t,x,y,w,r)\to \left(\lambda t,\lambda x, \lambda y,
\lambda^{\frac{2}{3}} w, \frac{r}{\lambda}\right),\label{scalingt}
\end{equation}
and therefore is expected to be holographically dual to
Lifshitz-like fixed points with space-like anisotropic scale
invariance. Note that the metric (\ref{scalingmetric}) is equivalent
to (\ref{aniso}) with $z=3/2$, $p=2$ and $d=3$.

By redefining the radius coordinate $\rho\equiv r^{\f{2}{3}}$ and
rescaling $(t,x,y,w)$ accordingly, we can rewrite the metric
(\ref{scalingmetric}) into another illuminating form \be
ds^2_E=\ti{R}^2\left[\rho^3(-dt^2+dx^2
+dy^2)+\rho^{2}dw^2+\f{d\rho^2}{\rho^2}\right] +R^2ds_{X_5}^2.
\label{scalingmetricz} \ee This can be regarded as gravity duals of
Lifshitz-like fixed points with $z=3/2$. It coincides with the metric
(\ref{AdSani}) with $p=2$ and $d=3$.

\subsection{Holographic Interpretation in terms of D3-D7 System}

Since our solution (\ref{scalingmetric}) is sourced by the RR 5-form
(\ref{fivef}) and 1-form flux (\ref{onef}), we expect it to be
interpreted as a D3-D7 system in string theory. When we compactify
the $w$ direction such that $w\sim w+L$ and place $N$ D3-brane along
the $(t,x,y,w)$ directions and $k$ D7-branes along the $(t,x,y,X_5)$
directions:
\begin{equation}\label{D3-D7}
\begin{array}{r|cccc|c|ccccccl}
\,\, \mbox{$\mathcal{M}_4\times S^1\times X_5$}\,\,\, & t     & x_1
& y & r  & w &  s_1   & s_2   & s_3   & s_4 & s_5 &\, \nonumber\\
\hline N \,\, \mbox{D3}\,\,\,& \x &   \x  &  \x &   & \x &   &  & &
&
&\,\nonumber\\
k\,\, \mbox{D7}\,\,\,& \x &   \x  &  \x &   &  &  \x & \x & \x &
\x&\x &\,\nonumber
\end{array}
\end{equation}
these $N$ D3 and $k$ D7 branes can source the desired RR 5-form and
1-form fluxes with
\begin{equation}
\ap=\f{(2\pi)^4N}{\mbox{Vol}(X_5)},\qquad \qquad \beta=\f{k}{L}.
\label{relationab}
\end{equation}
This brane configuration is the same as the one constructed to model
the fractional quantum Hall effect in \cite{FLRT}.

The number of the D3-branes determines the radii $R$ and $\tilde{R}$
in the scaling solution (\ref{scalingmetric}):
\begin{equation}
R^2=\f{12}{11}\ti{R}^2=2\sqrt{\frac{\pi^4}{\mbox{Vol}(X_5)}N}.
\end{equation}
For $X_5=S^5$ (whose volume is $\pi^3$),
$R^2=\f{12}{11}\ti{R}^2=2\sqrt{\pi N}$. The number of the D7-branes
gives the string coupling at $r=1$:
\begin{equation}
e^{\phi_0}=\frac{\sqrt{22}}{3}\frac{L}{k}.
\end{equation}

Now we say a few words about the field theory living on this D3-D7
system. We take $X_5=S^5$ to simplify the arguments. If we start
with $N$ D3-branes, whose low energy theory is the four-dimensional
${\cal N}=4$, $SU(N)$ super Yang-Mills theory, then the additional
$k$ D7-branes will source a non-trivial axion field
$\chi=\f{k}{L}w$, which in turn induces a $w$-dependent $\theta$
term (i.e. the topological term) of the Yang-Mills theory
\begin{equation}
\f{1}{4\pi}\int
\chi(w) \mbox{Tr} F\we F. \label{topym}
\end{equation}

For finite $\beta$ and $k$, $w$-direction is compactified. After
integrating over $w$, the 4D topological term (\ref{topym}) becomes
a 3D Chern-Simons term at level $k$:
\begin{equation}
\f{k}{4\pi}\int_{R^{1,2}} \mbox{Tr} \left[A\we F+\f{2}{3}A^3\right].
\end{equation}
Now we have two choices of the boundary condition for the
$w$-circle: periodic or anti-periodic. If we impose the
anti-periodic one, all fermions will become massive. This breaks all
supersymmetries and gives masses to scalar fields through quantum
corrections. In the IR limit, only a pure Yang-Mills term is left of
the original 4D $\mathcal{N}=4$ super Yang-Mills part of the action.
Since in the IR limit, the Chern-Simons term dominates this
Yang-Mills term, the final three-dimensional theory is a pure
Chern-Simons theory. In \cite{FLRT}, this D3-D7 system was
constructed to holographically model the FQHE precisely because it
flows to the pure Chern-Simons gauge theory in the IR. In this
model, the AdS/CFT correspondence in the IR limit manifests itself
as the level-rank duality of the pure Chern-Simons gauge theory.

On the other hand, if we take $k\rightarrow \infty$ (and
simultaneously $L\rightarrow \infty$) while keeping $\beta$ finite,
the $w$-direction is non-compact and the field theory is
four-dimensional. Even though the interaction (\ref{topym}) looks
non-local at the first sight, its contribution to the equations of
motion is actually local. This remarkable property occurs only when
$\chi(w)$ is a linear function of $w$ (as is the case here).

One might still doubt any relations of our new background (\ref{scalingmetric})
to the ${\cal N}=4$ super Yang-Mills theory as it is not
asymptotically $AdS_5$. One might also worry that the
dilaton (\ref{pdilaton}) blows up near the boundary $r\to \infty$.
However, as we will show in the next section, we can in fact
construct solutions which interpolate between the $AdS_5$ and our
scaling solution (\ref{scalingmetric}). This interpolating solution
can be considered as the dual of the RG flow between the two
systems. Notice that this caps off the strongly coupled region of
the scaling solution. We will also present anisotropic solutions for
analogous D4-D6 systems in section \ref{sec:dfds}.

\subsection{Black Brane Solutions and Entropy}

One more interesting fact about our scaling solutions is that we can
straightforwardly generalize them to black brane solutions which
have regular event horizons. The metric in the Einstein frame is
\begin{equation}
ds^2_E=\ti{R}^2\left[r^2(-F(r)dt^2+dx^2
+dy^2)+r^{\f{4}{3}}dw^2+\f{dr^2}{r^2F(r)}\right] +R^2ds_{X_5}^2,
\label{bhsol}
\end{equation}
where
\begin{equation}
F(r)=1-\f{\mu}{r^{\f{11}{3}}}.\label{frnon}
\end{equation}
The constant $\mu$ represents the mass parameter of the black brane.
The dilaton and RR fields remain the same.

Requiring the smoothness of the Euclidean geometry of (\ref{bhsol})
gives the Hawking temperature
\begin{equation}
T_H=\f{11}{12\pi}\mu^{\f{3}{11}}.
\end{equation}
The Bekenstein-Hawing entropy is then
\begin{equation}
S_{BH}=\gamma \cdot 
\left(\frac{\pi^3}{\mbox{Vol}(X_5)}\right)\cdot N^2 \cdot
T_H^{\f{8}{3}} \cdot V_2 \cdot L , \label{entbh}
\end{equation}
where $\gamma$ is a numerical factor
\begin{equation}
\gamma=2^{\f{4}{3}}\cdot 3^{\f{7}{6}}\cdot 11^{-\f{7}{6}}\cdot
\pi^{\f{5}{3}}\simeq 3.729
\end{equation} and $V_2$ represents the
area in the $(x,y)$ direction. The entropy (\ref{entbh}) is
proportional to $N^2$ and thus is consistent with the planar limit
of a certain gauge theory.

Notice that the power $8/3$ of temperature in (\ref{entbh}) can also be obtained from
a simple dimension counting. From the metric (\ref{bhsol}), the coordinate $w$ has the fractional
dimension $2/3$, while each of $(t,x,y)$ carries the unit dimension.

\section{RG Flow in AdS$_5$/CFT$_4$ and Scaling Solution}
\label{sec:RG}

In the previous section, we find a new scaling solution of the D3-D7
system in type IIB supergravity. To clarify its physical
interpretation, we will show below that we can construct
interpolating solutions that approach the $AdS_5\times X_5$
solutions in the $r\to \infty$ limit (i.e. UV limit of the
holographic duals) and the scaling solutions in the opposite limit
$r\to 0$. Then via the AdS/CFT correspondence, we can argue that the
system dual to our scaling solution is connected to the one dual to
the $AdS_5$ through the RG flow.

\subsection{Further Reduction of Equations of Motion}

To find the interpolating solution, we start with the general form
(\ref{ansatzmetric}). To simplify the problem we impose some extra
constraints which are consistent with both the $AdS_5$ and the
scaling solutions.

First, we can make the function $a(r)$ vanish by a reparametrization
of $r$:
\begin{equation}
a(r)=0. \label{constzero}
\end{equation}
Secondly, recall that we showed $[(2b-2a-\phi-2h)'e^{2z}]'=0$ for
generic solutions. In fact, both the $AdS_5$ and the scaling
solution satisfy a much stronger condition
\begin{equation}\label{constone0}
(2b-2a-\phi-2h)'=0.
\end{equation}
Since we are looking for a solution that interpolates between the
$AdS_5$ and the scaling solution, it is reasonable to impose
(\ref{constone0}) as a simplifying ansatz, namely
\begin{equation}
h(r)=b(r)-a(r)-\f{1}{2}\phi(r)+\ti{h}_0,\label{constone}
\end{equation}
where $\ti{h}_0$ is a constant. Similarly, since both the $AdS_5$
and the scaling solution have the nice property that in the Einstein
frame the radius of $S^5$ is a constant, we will also impose this
condition on our interpolating solution, namely we require
\begin{equation}
\phi(r)=4c(r)+4\log
r+\ti{\phi}_0,\label{consttwo}
\end{equation}
where $\ti{\phi}_0$ is a constant. These relations
(\ref{constzero}), (\ref{constone}) and (\ref{consttwo}) are the
constraints mentioned and are assumed throughout this section.

Under this ansatz, the Einstein frame metric becomes
\begin{equation}
ds^2_E=\f{e^{2b(r)-2c(r)-\frac{\tilde{\phi}_0}{2}}}{r^2}(-dt^2+dx^2+dy^2)+e^{-\f{3}{2}\ti{\phi}_0+2\ti{h}_0}
\f{e^{2b(r)-6c(r)}}{r^6}dw^2+e^{-\f{\ti{\phi}_0}{2}}\f{dr^2}{r^2}+e^{-\f{\ti{\phi}_0}{2}}ds^2_{X_5},\label{anmet}
\end{equation}
and the equations of motion are greatly simplified:
\ba
&& \ap=4e^{-\ti{\phi}_0},\\
&& b''=\f{2}{r^2}
+\f{23}{r}b'-10b'^2-\f{16c'}{r}+24b'c'-8c'^2, \label{eombdd}\\
&& c''=\f{4}{r^2}
+\f{14}{r}b'-6b'^2-\f{5c'}{r}+14b'c'-2c'^2, \label{eombddd}\\
&&
\f{1}{4}e^{-2\ti{h}_0+3\ti{\phi}_0-2b+14c}r^{14}\beta^2=6-8-6r^2b'^2-16rc'-8r^2c'^2+18rb'(1+rc').\label{eombd}
\ea We can confirm that the derivative of the r.h.s of (\ref{eombd})
is vanishing if (\ref{eombdd}), (\ref{eombddd}) and (\ref{eombd})
are satisfied. This means that the constraint (\ref{eombd}) is
consistent with (\ref{eombdd}) and (\ref{eombddd}).

\subsection{Interpolating Solution between $AdS_5$ and D3-D7 Scaling Solution}
\label{D3-D7Interpolating}

Now the problem amounts to solving the system of two coupled
first-ordered nonlinear ODEs (\ref{eombdd}) and (\ref{eombddd})
under the constraint (\ref{eombd}). First, notice that
(\ref{eombdd}) and (\ref{eombddd}) involve only the derivatives of
$b$ and $c$, thus once we find a solution to them, we can simply
choose the integration constants of $b(r)$ and $c(r)$ such that they
satisfy the constraint (\ref{eombd}) --- as long as it allows the
r.h.s of (\ref{eombd}) to be positive.\footnote{However as we will
show later this requirement is actually automatically satisfied by
the interpolating solution we are looking for; thus it does not
impose any additional constraint.} Therefore essentially we only
need to solve (\ref{eombdd}) and (\ref{eombddd}).

Next we redefine the radial coordinate $r$ and the derivatives of
the functions $b(r)$ and $c(r)$  as follows
\begin{equation}
s\equiv\log r,\ \ \ B(s)\equiv\f{\de b(r)}{\de \log r},\ \ \ \
C(s)\equiv\f{\de c(r)}{\de \log r}.
\end{equation}
Then the equations (\ref{eombdd}) and (\ref{eombddd}) are simply a
pair of first-ordered non-linear ODEs:
\begin{eqnarray}
&&\dot{B}=2+24B-16C-10B^2+24BC-8C^2,\label{eomsa}\\
&&\dot{C}=4+14B-4C-6B^2+14BC-2C^2\label{eomsb}.
\end{eqnarray}
where $\dot{B}\equiv \f{dB}{ds}$. A physical solution also needs to
satisfy \be (9B-8C-8)^2-(33B^2+48)<0, \label{region} \ee due to the
constraint (\ref{eombd}).

The dynamical system (\ref{eomsa}) and (\ref{eomsb}) has four fixed
points $(B,C)^{*}$ which can be classified into two pairs:
\begin{equation}
(B,C)^{*}=(\pm 1,- 1) \qquad \qquad \textrm{and} \qquad \qquad
(B,C)^{*}=\left(\pm \frac{7}{\sqrt{33}}, \pm \frac{1}{\sqrt{33}}-1\right).
\end{equation}
Inside each pair, the two fixed points are related by a coordinate
redefinition $r\rightarrow \frac{1}{r}$ thus are equivalent. The
fixed point
\begin{equation}\label{adsfive}
(B,C)^{*}_{AdS_5}=(1,- 1)
\end{equation}
corresponds the standard $AdS_5\times X_5$ solution.\footnote{The
fixed point $(B,C)^{*}=(-1,- 1)$ can be considered as its conjugate
since they are connected under a coordinate redefinition
$r\rightarrow \frac{1}{r}$. }

Our scaling D3-D7 solution (at zero temperature) corresponds to the
fixed point
\begin{equation}
(B,C)^{*}_{scaling}=\left(\f{7}{\s{33}},\f{1}{\s{33}}-1\right)\simeq
(1.2185,-0.8259).\label{scalefix}
\end{equation}
One can easily see that the metric (\ref{anmet}) with
(\ref{scalefix}) is equivalent to (\ref{scalingmetric}) via the
redefinition of radial coordinate $r\to r^{\s{33}/6}$. Since the two
fixed points with ``$-$" sign are equivalent to the two with ``$+$" sign
and are disconnected from them, we will not consider those any
further.

Now let's study the behavior of this dynamical system (\ref{eomsa})
and (\ref{eomsb}). Near the $AdS_5$ fixed point (\ref{adsfive}), the
eigensystem of the linear perturbations (defined by
$B(s)=B^{*}_{AdS_5}+\ep_b(s)$ and $C(s)=C^{*}_{AdS_5}+\ep_c(s)$ for
$\ep_b,\ep_c \ll 1$) is
\begin{equation}
\dot{\ep_b}= -20\ep_b+24\ep_c,\qquad \qquad \dot{\ep_c}=-12\ep_b+14\ep_c,
\end{equation} and both eigenvalues are negative: $\lambda_1=-4$ and
$\lambda_2=-2$; therefore the $AdS_5$ solution is a stable fixed
point as the system flows to the UV (i.e. $r\to \infty$).

On the other hand, the eigensystem of the linear perturbation near
the D3-D7 scaling fixed point (defined by
$B(s)=B^{*}_{scaling}+\eta_b(s)$ and
$C(s)=C^{*}_{scaling}+\eta_c(s)$ for $\eta_b,\eta_c \ll 1$) is \be
\dot{\eta_b}= -\f{116}{\s{33}}\eta_b+\f{152}{\s{33}}\eta_c,\ \ \ \
\dot{\eta_c}=-\f{70}{\s{33}}\eta_b+\f{94}{\s{33}}\eta_c. \ee In
contrast to the stable $AdS_5$ fixed point, this scaling solution
fixed point has one negative eigenvalue
($\lambda_1=-\f{\s{33}+\s{105}}{3}$) and one positive one
($\lambda_2=\f{\s{105}-\s{33}}{3}$). Therefore the fixed point
corresponding to the scaling solution is unstable. Near the
neighborhood of this scaling fixed point, there exist two special
trajectories: one corresponding to the negative eigenvalue
$\lambda_1$ and one to the positive $\lambda_2$. The fixed point
behaves like a UV (resp. IR) fixed point when approached along the
trajectory corresponding to the negative (resp. positive)
eigenvalue. When the fixed point is approached along a generic
direction, the trajectory only passes near it and then turns to flow
to infinity --- only one fine-tuned trajectory can reach the fixed
point.

\begin{figure}[htbp]
   \centering
   \subfigure[]{\label{fig:BCFlowCurveExact}
     \includegraphics[width=8cm]{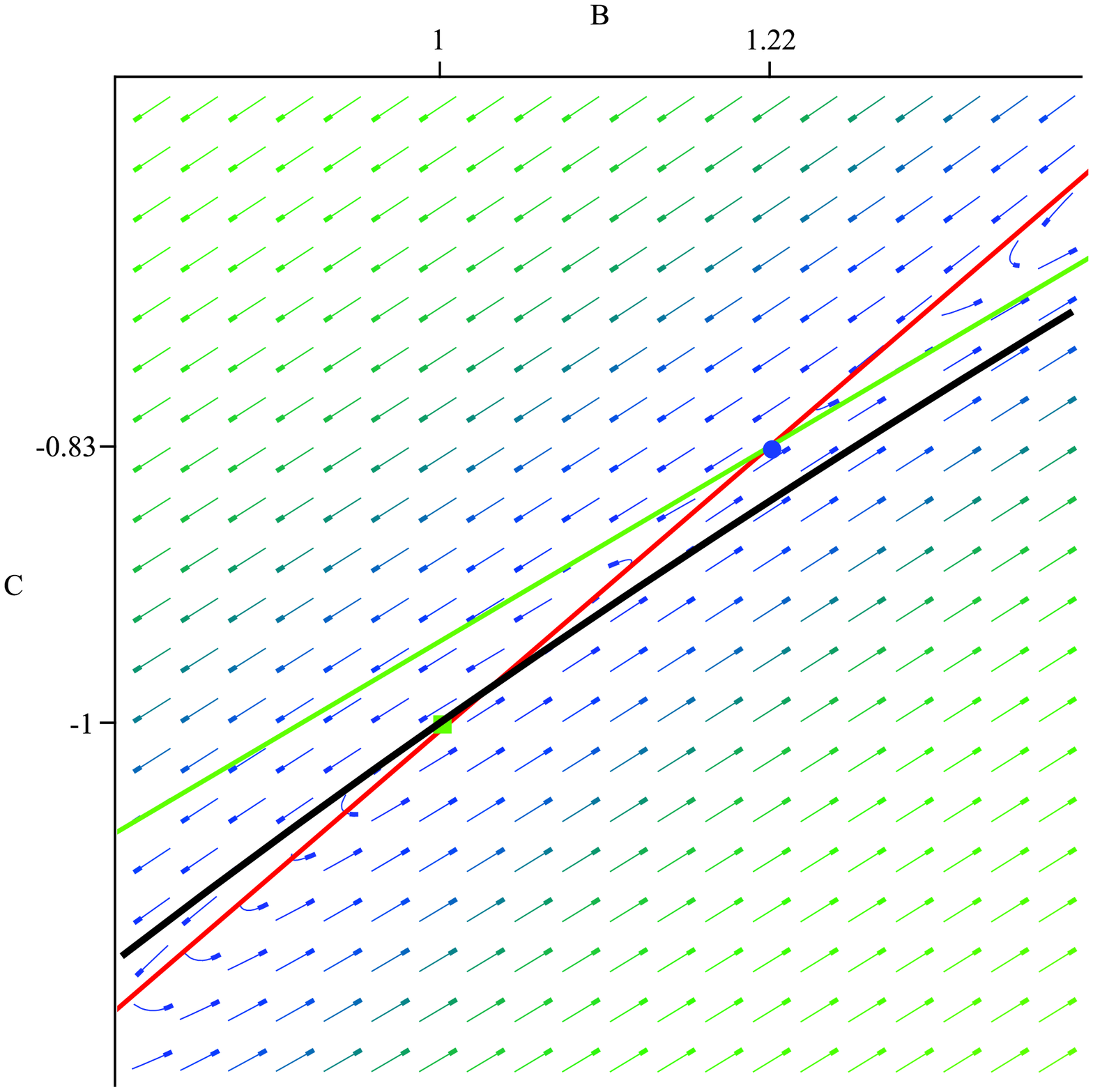}}
   \subfigure[]{\label{fig:flowdiagramHW}
     \includegraphics[height=6cm]{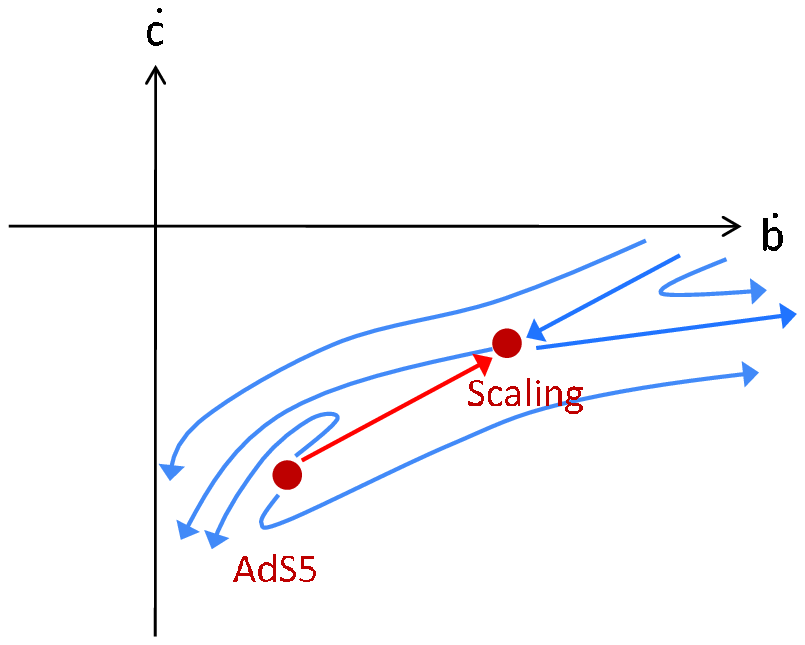}}
   \caption[]{The $(B,C)$ flow diagram. The horizontal and vertical axes
are $B\equiv\dot{b}$ and $C\equiv\dot{c}$, respectively. The arrows
point in the direction from the UV ($r=\infty$) to the IR ($r=0$).
In the left figure, the blue dot at
$(\frac{7}{\sqrt{33}},\frac{1}{\sqrt{33}}-1)\simeq (1.22,-0.83)$ is
the unstable fixed point corresponding to the scaling solution; the
green dot at $(1,-1)$ is the stable UV fixed point corresponding to
the $AdS_5$ solution. The green line running through the scaling
fixed point corresponds to the negative eigenvalue $\lambda_1$ while
the red line corresponds to the positive one $\lambda_2$. The black
curve is given by $9B-8C-8= \sqrt{33B^2+48}$ and corresponds to a
pure D3 solution; the allowed D3-D7 solutions are above this curve.}
\label{fig:FlowDiagram}
\end{figure}

Figure \ref{fig:FlowDiagram} shows the global behavior of this
dynamical system (\ref{eomsa}) and (\ref{eomsb}). Figure
\ref{fig:BCFlowCurveExact} is generated by numerical computations
and the salient features are schematically highlighted in the
hand-rendered Figure \ref{fig:flowdiagramHW}. The arrows point in
the direction from the UV ($s=\infty$) to the IR ($s=-\infty$). It
is easy to see from the direction fields that the $AdS_5$ fixed
point (the green dot at $(1,-1)$) is a stable UV fixed point while
the scaling fixed point (the blue dot at
$(\frac{7}{\sqrt{33}},\frac{1}{\sqrt{33}}-1)$) is unstable. Recall
that a physical solution needs to satisfy (\ref{region}). The
allowed region in the flow diagram are between two hyperbolic lines
given by $9B-8C-8=\pm \sqrt{33B^2+48}$. The black curve in Figure
\ref{fig:BCFlowCurveExact} is the one with the ``$+$" sign; and the
other one with the ``$-$" sign is its mirror image in the upper-left
corner but is out of the range of Figure \ref{fig:BCFlowCurveExact}.
It is clear that both the $AdS_5$ fixed point and the scaling
solution fixed point are in the allowed region. And since the
$AdS_5$ is a stable UV fixed point and there is no critical surface
separating it from the scaling solution fixed point, there exists a
trajectory emanating from the $AdS_5$ fixed point and flowing to the
scaling solution fixed point. Namely there exists a solution that
interpolates between the $AdS_5\times X_5$ solution in the UV
$(r\rightarrow \infty)$ and the D3-D7 scaling solution in the IR
$(r=0)$.

Now to solve the interpolating solution, we first choose the
integration constant. First, $\tilde{\phi}_0$ is determined only by
the 5-form flux:
$\tilde{\phi}_0=-\log{\frac{\alpha}{4}}=-\log{(\frac{4\pi^4}{\mbox{Vol}(X_5)}N)}$.
Then without loss of generality, we set
$\tilde{h}_0=\frac{\tilde{\phi}_0}{2}$ and choose the boundary
condition for $(b,c)$ to be
\begin{eqnarray}\label{bc}
&&b(r)\rightarrow \log{r}, \qquad c(r)\rightarrow -\log{r} \qquad
\textrm{at} \qquad r\rightarrow +\infty,\\
&&b(r)\rightarrow \frac{7}{\sqrt{33}}\log{r}+b_0,\qquad
c(r)\rightarrow \left(\frac{1}{\sqrt{33}}-1\right)\log{r}+b_0 \qquad
\textrm{at}\qquad r\rightarrow 0,
\end{eqnarray}
where $b_0\equiv
\frac{1}{6}\log{(\frac{4\pi^4}{\mbox{Vol}(X_5)}N)}-\frac{1}{12}\log{\frac{3}{8}}-\frac{1}{6}\log{
\beta}$. Then the dynamical system (\ref{eomsa}) and (\ref{eomsb})
plus the constraint
\begin{equation}\label{BCconstraint}
\frac{1}{8}[(9B-8C-8)^2-(33B^2+48)]=-\frac{\beta^2}{64\left(\frac{\pi^4}{\mbox{Vol}(X_5)}\right)^2N^2}e^{-2b(r)+14c(r)}r^{14}<0,
\end{equation}
and the boundary conditions (\ref{bc}) determine an interpolating
solution that approaches the $AdS_5\times X_5$ solution
\begin{eqnarray}
ds^2_E&=&R^2\left[r^2(-dt^2+dx^2+dy^2+dw^2)+\f{dr^2}{r^2}\right]+R^2ds^2_{X_5},\no
e^{\phi}&=&e^{\tilde{\phi}_0}, \label{AdSfive}
\end{eqnarray}
in the UV and the D3-D7 scaling solution
\begin{eqnarray}
ds^2_E&=&R^2[r^{\frac{12}{\sqrt{33}}}(-dt^2+dx^2+dy^2)
+\frac{dr^2}{r^{2}}]+\rho^2r^{\frac{8}{\sqrt{33}}}dw^2+R^2ds^2_{X_5},\\
e^{\phi}&=&e^{4b_0}e^{\tilde{\phi}_0}r^{\frac{4}{\sqrt{33}}},
\end{eqnarray}
in the IR. Here
$R^2=e^{-\tilde{\phi}_0/2}=2\sqrt{\frac{\pi^4}{\mbox{Vol}(X_5)} N}$
and $\rho^2=e^{-4b_0}R^2$. And the fluxes are
\begin{equation}
F_5=4R^4\left(\Omega_{X_5}+*\Omega_{X_5}\right), \qquad \qquad
F_1=\beta dw,\no
\end{equation}
throughout the system.

\begin{figure}[htbp]
   \centering
   \subfigure[$(B,C)$ of the interpolating solution.]{\label{fig:BC}
     \includegraphics[width=7cm]{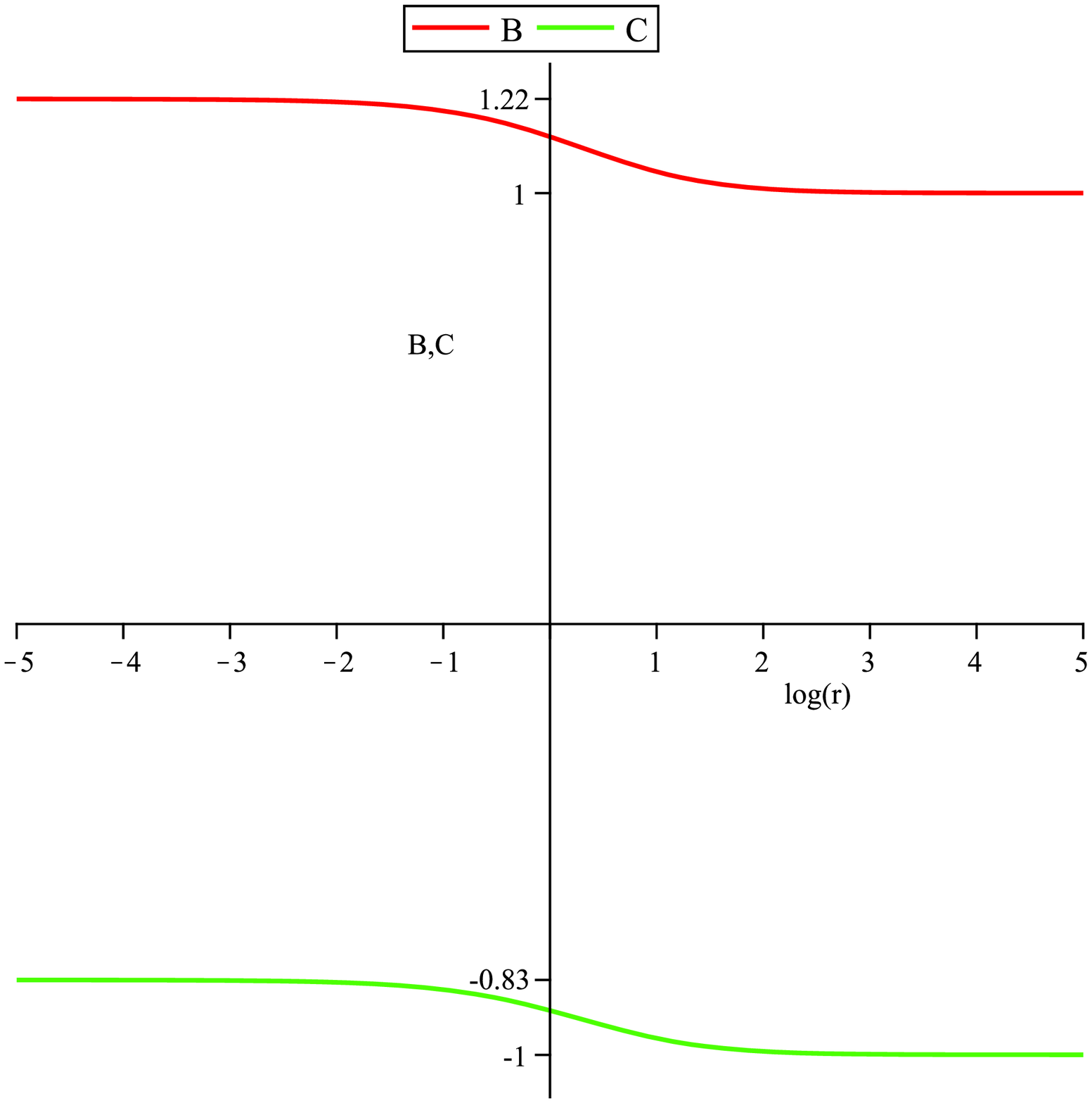}}
   \subfigure[Scalings of the interpolating solution in the Einstein Frame.]{\label{fig:BCPhi}
     \includegraphics[width=7cm]{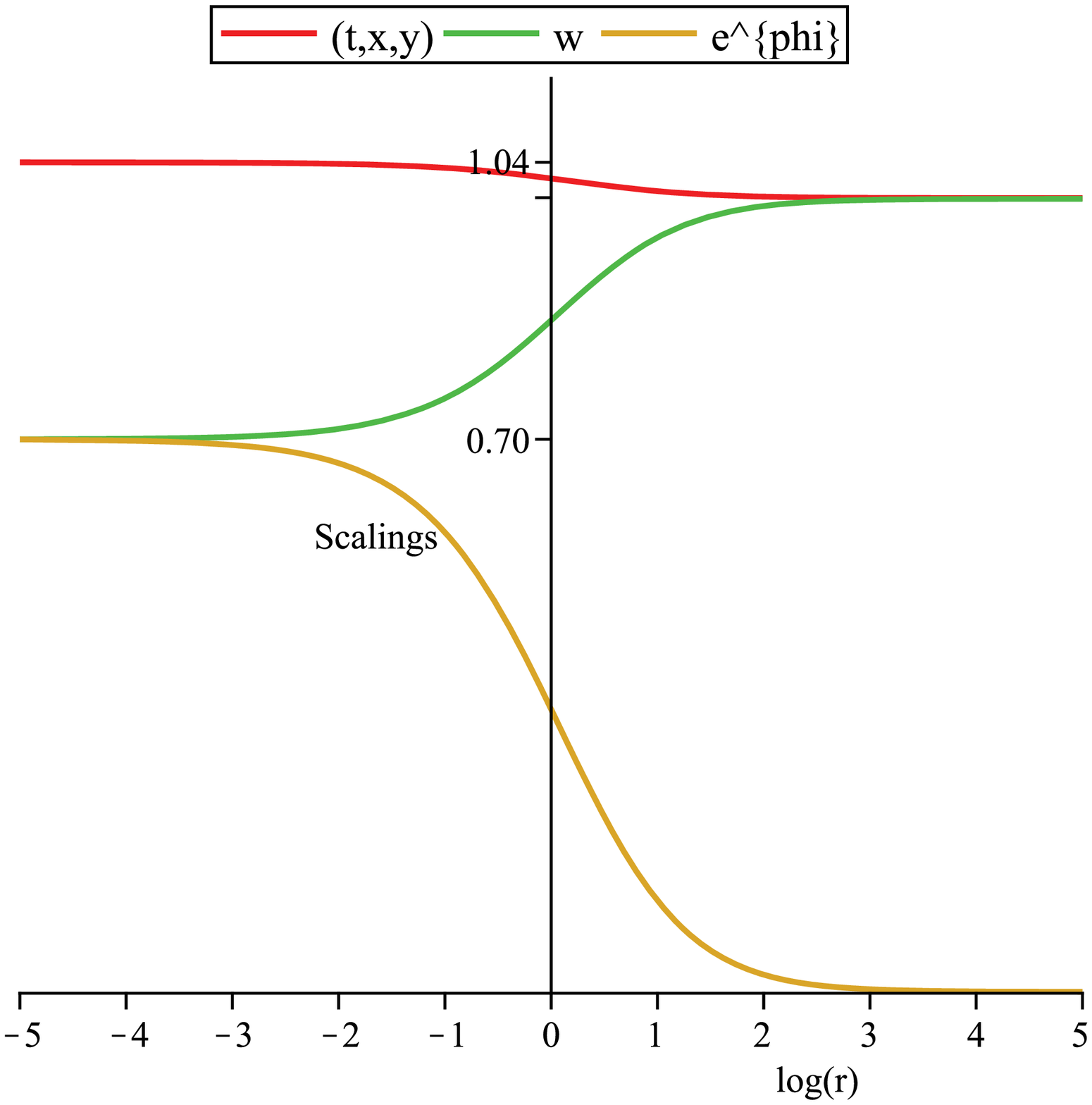}}
   \caption[]{An interpolating solution with fluxes satisfying
   $\frac{\beta^2}{64(\frac{\pi^4}{\mbox{Vol}(X_5)})^2N^2}=1$.
   In the left figure, $(B,C)\equiv (\dot{b},\dot{c})$ flow from
   $(\frac{7}{\sqrt{33}},\frac{1}{\sqrt{33}}-1)\simeq (1.22,-0.83)$
   in the IR to $(1,-1)$ in the UV. The right figure shows that,
   in the Einstein frame, the scalings of
   the $(t,x,y)$-directions, the $w$-direction, and
   $e^{\phi}$ flow from
   $(\frac{8}{\sqrt{33}},\frac{4}{\sqrt{33}},\frac{4}{\sqrt{33}})
   \simeq (1.04,0.70,0.70)$ in the IR to $(1,1,0)$ in the UV.}
   \label{fig:BCPhiInterpolating}
   \end{figure}

We can easily solve $(b,c)$ numerically for arbitrary fluxes.
Systems with different flux numbers differ only in their speeds in
approaching the fixed points. Figure \ref{fig:BCPhiInterpolating}
shows the system with
$\frac{\beta^2}{64(\frac{\pi^4}{\mbox{Vol}(X_5)})^2N^2}=1$ as an
example. Figure \ref{fig:BC} graphs the behavior of
$(\dot{b},\dot{c})$, and more directly, Figure \ref{fig:BCPhi}
presents the scalings of the $(t,x,y)$-directions, $w$-direction,
and $e^{\phi}$ in the Einstein frame as the interpolating solution
flows from the D3-D7 scaling solution in the IR $(r=0)$ to the
$AdS_5$ in the UV $(r\rightarrow \infty)$.

\subsection{Interpolating Solutions as Holographic RG Flows}

Here we try to interpret our interpolating solutions via AdS/CFT.
To make the argument simple we focus on $X_5=S^5$. Consider the
standard AdS$_5$/CFT$_4$ for the $\mathcal{N}=4$ super Yang-Mills. We
can perturb the $\mathcal{N}=4$ super Yang-Mills by many relevant
operators ${\cal O}_1, {\cal O}_2, \ddd$ as $\delta S=\int dx^4 [g_1
O_1+g_2 O_2+\ddd]$.

Our solution which flows from the $AdS_5\times S^5$ to the scaling
D3-D7 solution is dual to an anisotropic RG flow triggered by a
non-supersymmetric relevant deformation (\ref{topym}), called $O_1$.
This RG-flow eventually ends at the IR fixed point that is dual to
the scaling solution. Therefore $O_1$ becomes irrelevant in the IR
limit. However, our sketch of the holographic RG flow (see Fig.
\ref{fig:FlowDiagram}) tells us that this flow is unstable and
signals the presence of another relevant operator $O_2$ which
becomes relevant even at the IR fixed point. So if we slightly
perturb this flow by $O_2$, then the RG flow passes near the IR
fixed point and eventually goes to infinity. Its asymptotic behavior
will be derived analytically in the next subsection. However, we can
still fine tune such that there is no $O_2$ generated (i.e. $g_2=0$)
to realize the IR fixed point governed by the scaling solution. This
is exactly what experimentalists usually do to realize an unstable
fixed point.\footnote{It is feasible because as long as the starting
point is close enough to the critical line (which corresponds to the
fine-tuned trajectory that hits the fixed point), the trajectory
will stay for a very long time near the fixed point to allow the
measurements \cite{Cardy}.}

A similar situation occurs when we perturb the Heisenberg model by
an anisotropic Ising-like interaction: $ H=-J\sum_{\la ij\lb
}(\sigma^x_i\sigma^x_j+\sigma^y_i\sigma^y_j
+\sigma^z_i\sigma^z_j)-D\sum_{\la ij\lb }\sigma^z_i\sigma^z_j,$
where $D>0$. In this analogy, the UV fixed point is the Heisenberg
model fixed point and the IR one is the Ising model fixed point.
This structure of the RG flow is rather generic and is one example
of the phenomenon called crossover \cite{Cardy}.

\subsection{Exact Solutions without D7-brane Charge}

In our interpolating solution, the flow from the UV $AdS_5$ fixed
point reaches the IR D3-D7 scaling solution fixed point only if it
starts along a specific direction. One wonders where the flow would
end up if its initial direction slightly deviates from the
fine-tuned one. A closer look at Figure \ref{fig:BCFlowCurveExact}
tells us that if it deviates from the desired direction slightly to
the left, it will turn back before reaching the scaling solution
fixed point and asymptote to the $B<1$ part of the black curve given
by $9B-8C-8= \sqrt{33B^2+48}$ (due to the constraint
(\ref{region})). On the other hand, if it deviates slightly to the
right, it will pass near (but not hitting) the scaling solution
fixed point and then bend slightly downward and finally asymptote to
the $B>1$ part of the same black curve.

The curve given by $9B-8C-8= \sqrt{33B^2+48}$ is the solution
interpolating between the $AdS_5$ in the UV and some other solution
in the IR. Since it saturates the inequality (\ref{BCconstraint}),
there is no D7 brane charge ($\beta=0$): it is a solution of the
pure D3-brane system. Not only is it important because it gives the
asymptotic form of what our interpolating solution decays into when
perturbed by a relevant operator, it is also interesting as a pure
D3-brane solution other than $AdS_5$. In this subsection, we will
solve it analytically.

First, eliminating $C$ from equations (\ref{eomsa}) and
(\ref{eomsb}) gives \be
\f{dB}{ds}=4+\f{11}{4}B^2-\f{3}{4}B\s{48+33B^2}, \ee which can
actually be solved analytically. There are two solutions
distinguished by the $\pm$ sign: \ba &&B_{\pm}(s)=\f{e^{8s}\mp
6\s{\f{3}{11}}e^{4s}+1}{e^{8s}-1},\no &&C_{\pm}(s)=-\f{e^{8s}\pm
4\s{\f{3}{11}}e^{4s}-1}{e^{8s}-1}, \ea which lead to \ba &&
b_{\pm}(s)=\ti{b}_0-s\pm\f{3}{4}\s{\f{3}{11}}\log\f{e^{4s}+1}{e^{4s}-1}+\f{1}{4}\log(e^{8s}-1),\no
&&
c_{\pm}(s)=\ti{c}_0-s\pm\f{1}{2}\s{\f{3}{11}}\log\f{e^{4s}+1}{e^{4s}-1}.
\ea We can choose $\tilde{h}_0=\frac{\tilde{\phi}_0}{2}$ and
$\tilde{b}_0=\tilde{c}_0=0$; then the metric in the Einstein frame
and the dilaton are simply
\begin{eqnarray}\label{solexb}
ds^2_E&=&R^2\Bigg[
\sqrt{\frac{r^8-1}{r^4}}
\left(\frac{r^4+1}{r^4-1}\right)^{\pm\frac{1}{2}\sqrt{\frac{3}{11}}}
(-dt^2+dx^2+dy^2)+\sqrt{\frac{r^8-1}{r^4}}\left(\frac{r^4+1}{r^4-1}\right)^{\mp\frac{3}{2}\sqrt{\frac{3}{11}}}
dw^2 \non\\
&&\qquad+\frac{dr^2}{r^2}+ds^2_{X_5}\Bigg],\non\\
e^{\phi}&=&\left(\frac{r^4+1}{r^4-1}\right)^{\pm
2\sqrt{\frac{3}{11}}}e^{\tilde{\phi}_0},
\end{eqnarray}
with $R^2=2\sqrt{\frac{\pi^4}{\mbox{Vol}(X_5)} N}$.

Both solutions in (\ref{solexb}) become $AdS_5\times X_5$ in the UV
fixed point ($r\to \infty)$; however when going towards the IR, they
become singular at $r=1$. The solution with $``+"$ (resp. $``-"$)
sign covers $B>1$ (resp. $B<1$) part of the curve $9B-8C-8=
\sqrt{33B^2+48}$. Even though these zero temperature solutions
become singular ar $r=1$, their corresponding black brane solutions
at finite temperature are expected to be smooth. An explicit
construction of such black brane solutions is left as a future
problem. Also it would be very interesting to understand what
anisotropic relevant deformations of the ${\cal N}=4$ Yang-Mills are
dual to this background.

\section{Hydrodynamics}
\label{sec:viscosity}

As we obtained the black brane solutions \eqref{bhsol}, we can
consider the hydrodynamic behavior of their dual field theories from
the supergravity side \cite{PSS}. In this section, we especially
determine the shear and  bulk viscosities from dispersion relations
for the corresponding quasinormal modes of fluctuations around the
background \eqref{bhsol}, following \cite{KS, MT}.

For this purpose, we first reduce $S^5$ part and consider
fluctuations around the resulting five-dimensional background. The
procedure of the reduction and derivation of the linearized
equations of motion for the fluctuations are summarized in Appendix
\ref{zeromodeh}. In these equations, fluctuation of the resulting
five-dimensional metric $\delta g_{\mu\nu}^{(5)}=H_{\mu\nu}$,
dilaton $\delta \phi =\varphi$, axion $\delta \chi=\eta$, the trace
part of the $S^5$ metric $\pi$ and the five-form flux $\delta F_5=
f_5$ appear generally.

In this section, we assume that the $w$ direction is compactified
and the neglect the momentum in this direction so that we extract the effective
$2+1$ dimensional holographic dual theories from the total $3+1$ dimensional ones.
In this situation we can choose the momentum in $y$ direction.
 By considering the symmetry of the
background, we can decompose the fluctuation into channels decoupled
from each other in the linear order:
\begin{eqnarray}
\text{shear channel} && \qquad H_{tx}, H_{xy}, H_{xr}, \label{shear_ch}\\
\text{scalar channel} && \qquad H_{xw}, \\
\text{sound channel} &&\qquad H_{tt}, H_{ty}, H_{tr}, H_{xx},
H_{yy}, H_{yr}, H_{ww}, H_{rr}, \varphi,\pi,
f_{txywr},  \label{sound_ch}\\
\text{``axion channel"} && \qquad H_{tw}, H_{yw}, H_{rw}, \eta.
\end{eqnarray}
In this paper, we consider the shear and sound channel only, since
we can read off the shear and bulk viscosity from dispersion
relations for the quasinormal modes of these channels.

Before deriving the dispersion relation, we need to consider the
gauge fixing of the fluctuations. Let us assume that the
fluctuations are of the form
\begin{eqnarray}
 H_{\mu\nu} &=& \tilde{H}_{\mu\nu}(r)e^{-i(\omega t-qy)},  \qquad
 \varphi = \varphi(r)e^{-i(\omega t-qy)}, \nonumber \\
 \eta &=& \eta(r) e^{-i(\omega t-qy)},  \qquad\quad
 \pi = \pi (r) e^{-i(\omega t-qy)}, \\
 f_{\mu\nu\rho\sigma\lambda} &=& f_{\mu\nu\rho\sigma\lambda}(r) e^{-i(\omega t-qy)}. \nonumber
\end{eqnarray}
Especially, for later convenience, we rewrite some of their components as
\begin{eqnarray}
\tilde{H}_{tt}(r) &=& \tilde{R^2}c_{t}^2H_{tt}(r), \qquad
\tilde{H}_{ab}(r) = \tilde{R^2}c_{x}^2 H_{ab}(r), \\
\tilde{H}_{ta}(r) &=& \tilde{R^2}c_{x}^2 H_{ta}(r), \qquad
\tilde{H}_{ww}(r) = \tilde{R^2}c_{x}^{4/3} H_{ww}(r),
\end{eqnarray}
where $a,b=x$ or $y$, and $c_{t}^2=r^2F(r)$ and $c_{x}^2=r^2$; the function
$F(r)$ is defined by (\ref{frnon}). As in
Appendix \ref{zeromodeh}, we then take the radial gauge $H_{\mu
r}=0$ in the following discussion. Here we notice that there still
exist residual gauge degrees of freedom under an infinitesimal
diffeomorphism $x^{\mu}\to x^{\mu}+\xi^{\mu}$ with 
$\xi^{\mu}=\xi^{\mu}(r)e^{-i(\omega t-qy)}$:
\begin{eqnarray}
&&H_{\mu\nu}\to H_{\mu\nu}
-\nabla_{\mu}\xi_{\nu}-\nabla_{\nu}\xi_{\mu}, \qquad
\varphi \to \varphi -\xi_{\mu}\partial^{\mu}\phi, \\
&&\eta \to \eta -\xi_{\mu}\partial^{\mu}\chi,
\qquad \pi \to \pi-\xi_{\mu}\partial^{\mu}g_{\alpha}^{\ \alpha}\ .
\end{eqnarray}
Here the covariant derivative is defined by the background metric.
In \cite{KS}, it has been noticed that one can derive the dispersion
relation for the quasinormal modes by defining gauge invariant
quantities, instead of performing the gauge fixing completely.
Following this, we define the gauge invariant quantities as
\begin{eqnarray}
Z_{1}(r) =qH_{tx}(r)+\omega H_{xy}(r),
\end{eqnarray}
for the shear channel and
\begin{eqnarray}
&&Z_{0}(r) = q^2\frac{c_t^2}{c_x^2}H_{tt}(r) + 2q\omega H_{ty}(r)+\omega^2
H_{yy}(r)
+ \frac{3}{5}\left(q^2\frac{c_t'c_t}{c_x'c_x}-\omega^2\right)
(H_{xx}(r)+H_{ww}(r)), \nonumber\\
\\
&&Z_{\varphi}(r) = \varphi(r)-\frac{1}{2}H_{ww}(r), \\
&&Z_{\pi}(r) = \pi(r),
\end{eqnarray}
for the sound channel. Below, we derive equations for these
quantities by using linearized equations of motion summarized in
Appendix \ref{sec:viscosity_equation} and then the dispersion
relations for the quasinormal modes by imposing an appropriate
boundary condition.

\subsection{Shear Viscosity}
\label{subsec:shear_viscosity}

From the equations \eqref{shear1},\eqref{shear2} and \eqref{shear3},
we obtain the equation for $Z_1$ as
\begin{eqnarray}
Z_1'' +\frac{14F(r)(\mathfrak{q}^2F(r)-\mathfrak{w}^2)
-3\mathfrak{w}^2rF'(r)}{3rF(r)(\mathfrak{q}^2F(r)-\mathfrak{w}^2)}Z_1'
+\frac{121\mu^{\frac{6}{11}}(\mathfrak{w}^2-F(r)\mathfrak{q}^2)}{36r^4(F(r))^2}Z_1=0,
\end{eqnarray}
where the dimensionless frequency $\mathfrak{w}$ and momentum
$\mathfrak{q}$ are defined by
\begin{eqnarray}
\mathfrak{w} = \frac{w}{2\pi T_H}, \quad \mathfrak{q} =
\frac{q}{2\pi T_H},
\end{eqnarray}
respectively. 

In order to solve this equation, we consider the
asymptotic behavior of $Z_1$ first. As $r\to \infty$, 
$Z_1\sim Ar^{0}+Br^{-\frac{11}{3}}$ with constants $A$, $B$ 
generally. For the quasinormal modes, 
$A$ is set to zero, or, in other words, we should impose the 
Dirichlet boundary condition $Z_1=0$ at $r=\infty$. 
As $r\to \mu^{\frac{3}{11}}$,
$Z_1\sim (F(r))^{\pm i\mathfrak{w}/2}$ and, in order to ensure that only
incoming waves exist at the horizon, we take the one with negative sign. 

By taking the hydrodynamic
limit $\mathfrak{q}\ll 1$ and $\mathfrak{w}\ll 1$, we can
perturbatively determine $Z_1$ as
\begin{eqnarray}
Z_1=C(F(r))^{-i\mathfrak{w}/2}\left(1+\frac{i\mathfrak{q}^2}{2\mathfrak{w}}F(r)+\mathcal{O}(\mathfrak{q}^2,\mathfrak{w})\right),
\end{eqnarray}
where $C$ is a constant. Here we assumed that $\mathfrak{q}^2$ and
$\mathfrak{w}$ are of the same order as usual for the shear channel.
Then by imposing the Dirichlet condition $Z_1=0$ at $r = \infty$, 
we find $\mathfrak{w}
=-\frac{i\mathfrak{q}^2}{2}$. Comparing with the hydrodynamic
relation for the shear channel
\begin{eqnarray}
\omega = -\frac{i\eta}{T_H s}q^2,
\end{eqnarray}
we can find that the universal bound for the shear viscosity $\eta$
to the entropy density $s$ ratio \cite{KSS} is saturated for the
current case:
\begin{eqnarray}
\frac{\eta}{s} =\frac{1}{4\pi}.
\end{eqnarray}

\subsection{Bulk Viscosity}
\label{subsec:bulk_viscosity}

From the equations derived in Appendix \ref{subsec:sound_eq}, we
obtain the differential equations for $Z_{\varphi}$, $Z_{\pi}$ and
$Z_{0}$ as
\begin{eqnarray}
&&Z_{\varphi}''+\ln'(c_t^2c_x^{8/3})Z_{\varphi}' +\frac{1}{c_{t}^2}
\left(\frac{\omega^2}{c_{t}^2}-\frac{q^2}{c_{x}^2}\right)Z_{\varphi}
\nonumber \\
&&\qquad\qquad\qquad\qquad\qquad -\frac{44}{9c_t^2}Z_{\varphi}+
\frac{11(\mu^3-2\mu^2r^{11/3}+2r^{11})}{27r^{28/3}(r^{11/3}-\mu)}Z_{\pi}=0,
\label{z_varphi}\\
&&Z_{\pi}''+\ln'(c_t^2c_x^{8/3})Z_{\pi}' +\frac{1}{c_{t}^2}
\left(\frac{\omega^2}{c_{t}^2}-\frac{q^2}{c_{x}^2}\right)Z_{\pi}
-\frac{32}{c_t^2}Z_{\pi}=0, \label{z_pi}\\
&&Z_{0}''+\mathcal{F}(r)Z_{0}'+\mathcal{G}(r)Z_{0}+\mathcal{H}(r)Z_{\varphi}=0,
\label{z_0}
\end{eqnarray}
where
\begin{eqnarray}
&&\mathcal{F}(r) =
\frac{625\mu^3q^4-3584(q^2-\omega^2)^2r^{11}}{3r(\mu-r^{11/3})(5\mu
q^2-16r^{11/3}(q^2-\omega^2))^2}
\nonumber \\
&&\quad\quad\quad+\frac{-20\mu^2q^2(157q^2-112\omega^2)r^{11/3}
+32\mu(149q^2-24\omega^2)(q^2-\omega^2)r^{22/3}}{3r(\mu-r^{11/3})(5\mu q^2-16r^{11/3}(q^2-\omega^2))^2},\\
&&\mathcal{G}(r) =
\frac{3025\mu^4q^4-5\mu^2q^2r^{16/3}(333q^2-1936r^2)(q^2-\omega^2)-2304r^{38/3}(q^2-\omega^2)^3}
{9r^2(\mu-r^{11/3})^2(5\mu q^2-16r^{11/3}(q^2-\omega^2))^2}  \nonumber\\
&&\qquad\quad+\frac{3744\mu r^{9}q^2(q^2-\omega^2)^2
+5\mu^3q^2r^{5/3}(45q^4-2541q^2r^2+1936r^2\omega^2)}{9r^2(\mu-r^{11/3})^2(5\mu
q^2-16r^{11/3}(q^2-\omega^2))^2}
,\\
&&\mathcal{H}(r)= -\frac{22(1536r^{11}(q^2-\omega^2)^3-1440\mu
r^{22/3} q^2(q^2-\omega^2)^2)}
{45r^2(\mu-r^{11/3})(5\mu q^2-16r^{11/3}(q^2-\omega^2))^2} \nonumber \\
&&\qquad\qquad-\frac{22(-\mu^3(425q^2-550\omega^2)q^4+20\mu
^2q^2r^{11/3}(83q^4-171q^2\omega^2+88\omega^4))}
{45r^2(\mu-r^{11/3})(5\mu q^2-16r^{11/3}(q^2-\omega^2))^2}.
\end{eqnarray}
Then, as in the case of the shear channel, all we have to do next is
to solve these equations by imposing the incoming wave boundary
condition at the horizon and the Dirichlet boundary condition at
infinity to derive the dispersion relation for the quasinormal mode
of the sound channel. We also take the hydrodynamic limit and assume
that $\mathfrak{q}$ and $\mathfrak{w}$ are of the same order, as is
expected for the sound channel. For $Z_{\pi}=0$, we can see that the
nonsingular solution for \eqref{z_pi} is a constant, which turns out
to be zero due to the Dirichlet boundary condition $Z_{\pi}=0$ at
the boundary $r=\infty$. By substituting this into \eqref{z_varphi},
we obtain $Z_{\varphi}=0$ in a similar manner. Then from these
results and \eqref{z_0}, by using a similar ansatz $Z_{0} =
(F(r))^{-i\mathfrak{w}/2}Y_{0}(r)$, we can perturbatively determine
$Z_{0}$ as
\begin{eqnarray}
Z_{0} = \tilde{C}(F(r))^{-i\mathfrak{w}/2}
\left(1-\frac{5(1+2i\mathfrak{w})\mathfrak{q}^2}{11\mathfrak{q}^2-16\mathfrak{w}^2}F(r)+
\mathcal{O}(\mathfrak{q^2},\mathfrak{w}^2,\mathfrak{qw})\right),
\end{eqnarray}
where $\tilde{C}$ is a normalization constant. As a result of the
Dirichlet boundary condition $Z_{0}=0$ at $r=\infty$, we obtain the
dispersion relation for the sound channel
\begin{eqnarray}
\mathfrak{w} =
\frac{1}{2}\sqrt{\frac{3}{2}}\mathfrak{q}-i\frac{5}{16}\mathfrak{q}^2+\cdots.
\label{sound_dispersion_sugra}
\end{eqnarray}
Let us recall a hydrodynamic relation for the sound channel in the
noncompact $(d+1)$-dimensional spacetime
\begin{eqnarray}
\omega=c_sq-i\frac{\eta}{T_H
s}\left(\frac{d-1}{d}+\frac{\zeta}{2\eta}\right)q^2+\cdots,
\label{sound_dispersion_hydro}
\end{eqnarray}
where $c_s$ and $\zeta$ are the sound velocity and the bulk
viscosity, respectively. Since there are two noncompact spatial
dimensions for the dual field theory now,  $d=2$ for the current
case. Therefore, by comparing \eqref{sound_dispersion_sugra} with
\eqref{sound_dispersion_hydro} with $d=2$, we obtain
\begin{eqnarray}
c_s^2 = \frac{3}{8},\qquad \frac{\zeta}{\eta} = \frac{1}{4}.
\end{eqnarray}
As for $\zeta/\eta$ and $c_s$ of strongly coupled gauge theory
plasma in $d$ noncompact spatial dimensions, it is conjectured in
\cite{Buchel} that they satisfy an inequality
\begin{eqnarray}
\frac{\zeta}{\eta}\ge 2\left(\frac{1}{d}-c_s^2\right).
\end{eqnarray}
In our case, this inequality is saturated.

\section{Entanglement Entropy of D3-D7 Scaling Solutions}
\label{sec:EE}

When a quantum system is divided into two subsystems: $A$ and its
complement $B$, the von Neumann entropy $S_A=-\textrm{Tr} \rho_A
\log{\rho_A}$ (where $\rho_A$ is the reduced density matrix after
tracing out $B$) is called the entanglement entropy. The scaling
behaviors and certain universal\footnote{Here ``universal" means the
independence from the choice of different delineations of
subsystems.} coefficients of the entanglement entropy encode
important information on the degrees of freedom and non-local
correlations of the system \cite{BoSr,HoCa}.

For an anisotropic system, an interesting question is ``how does the
scaling behavior of the entanglement entropy depend on the direction
along which the subsystems are delineated?" In this section, we will
study the entanglement entropy of various subsystems $A$ of the
$(x,y,w)$ space at the boundary $(r\rightarrow \infty)$ of the 5D
part of the D3-D7 scaling solution (\ref{scalingmetric}). The field
theoretical computation of the entanglement entropy is expected to be
difficult as the system will be strongly coupled.
We will instead compute its holographic dual on the
gravity side. The holographic dual of the entanglement entropy of a
subsystem $A$ is given by
\begin{equation}
S_{EE}=\frac{\mbox{Area}^{min}}{4G_{5}},
\end{equation}
where $\mbox{Area}^{min}$ is the area of the three-dimensional
\emph{minimal surface} that lives inside the $(r,x,y,w)$ space and
borders on the boundary $\partial A$ of the subsystem $A$ \cite{RT}.

After a coordinate transformation $r=\frac{1}{z}$, the metric of the
D3-D7 scaling solution (\ref{scalingmetric}) becomes
\begin{equation}
ds^2=\tilde{R}^2
\left(\frac{-dt^2+dx^2+dy^2+dz^2}{z^{2}}+\frac{dw^2}{z^{4/3}}\right)+R^2ds^2_{X_5}.
\end{equation}
For $X_5=S^5$, $R^2=2\sqrt{\pi N}$ and
$\tilde{R}^2=\frac{11}{12}R^2$. We consider the full boundary system
given by
\begin{equation}
x\in [0,L_x], \qquad y\in [0,L_y],\qquad w \in [0,L_{w}].
\end{equation}
Among the
various types of subsystems, we will only study the easiest types:
stripes with either $x$ or (inequivalently) $w$ restricted to a
smaller length.

\subsection{Entanglement Entropy for Subsystem along $x$-direction}
Let's first consider the subsystem $A$ cut out along the
x-direction:
\begin{equation}
x\in [0,\ell_x< L_x], \qquad y\in [0,L_y],\qquad w \in [0,L_{w}].
\end{equation}
The three-dimensional minimal surface bordering on $\partial A$ is
given by the embedding function $z=z(x)$:
\begin{equation}
\mbox{Area}^{min}_{x}=\int^{\ell_x}_{0}dx\int^{L_y}_{0}dy
\int^{L_{w}}_{0}dw \frac{\tilde{R}}{z}\cdot
\frac{\tilde{R}}{z^{2/3}}\cdot
\sqrt{\left(\frac{\tilde{R}}{z}\right)^2+\left(\frac{\tilde{R}}{z}z'\right)^2}
=\tilde{R}^3L_yL_{w}\int^{\ell_x}_{0}dx\frac{1}{z^{d}}\sqrt{1+z'^2},
\end{equation}
where $d=1+1+\frac{2}{3}=\frac{8}{3}$ is the total scaling of the
boundary system.

This is a Lagrangian system with
$\mathcal{L}=\tilde{R}^3L_yL_{w}\frac{1}{z^{d}}\sqrt{1+z'^2}$. The
$z(x)$ that minimizes the surface area is then given by the equation
of motion
\begin{equation}
z'=\pm\sqrt{\left(\frac{z_{*}}{z}\right)^{2d}-1},
\end{equation}
where $z_{*}$ is the peak of $z$ on the minimal surface, across
which $z'$ changes sign. It can be solved from $
\ell_x=2\int^{z_{*}}_{0}\frac{dz}{z'}$
, which gives
$z_{*}=\frac{\ell_x}{2}\frac{\Gamma(\frac{1}{2d})}{\sqrt{\pi}\Gamma(\frac{1}{2d}+\frac{1}{2})}
$. The minimal surface is then
\begin{equation}
\mbox{Area}^{min}_{x}=2\tilde{R}^3L_yL_{w}\frac{1}{z^{d-1}_{*}}\cdot In
\qquad \textrm{with} \qquad
In\equiv\int^{1}_{0}du\frac{1}{u^{d}}\frac{1}{\sqrt{1-u^{2d}}}.
\end{equation}

$In$ has a UV divergence at $z\rightarrow 0$. Imposing the UV cutoff
by choosing the lattice spacing $a$ for the boundary system, we get
\begin{equation}
In=\frac{1}{d-1}\left(\frac{1}{(a/z_{*})^{d-1}}-\frac{\sqrt{\pi}\Gamma(\frac{1}{2d}+\frac{1}{2})}{\Gamma(\frac{1}{2d})}\right).
\end{equation}
Then plugging the value of $z_{*}$ and the five-dimensional Newton's
constant $G_{5}=\frac{G_{10}}{V_{X_5}}$ with $G_{10}=8 \pi^6
\ell^8_s$, we finally obtain the holographic entanglement entropy
for the subsystem divided out along the x-direction:
\begin{eqnarray}
S_{EE-x}&=&\left(\frac{11}{12}\right)^3\frac{\pi^2}{\mbox{Vol}(X_5)}\cdot N^2
L_yL_{w}\frac{1}{d-1}\left[\frac{1}{a^{d-1}}-\left(\frac{2}{\ell_x}\right)^{d-1}\left(\frac{\sqrt{\pi}
\Gamma(\frac{1+d}{2d})}{\Gamma(\frac{1}{2d})}\right)^{d}\right]\non\\
&=& N^2
L_yL_{w}\left[\frac{\gamma_1}{a^{5/3}}-\frac{\gamma_2}{(\ell_x)^{5/3}}\right],
\end{eqnarray}
with $d=\frac{8}{3}$. $\gamma_1$ and $\gamma_2$ are numerical
constants.

Now let's interpret the result. First, the holographic entanglement
entropy is proportional to the area of the boundary of the subsystem
$\partial A = L_yL_w$ --- as expected from the ``area law" \cite{BoSr}
for the entanglement entropy from direct field theory computations. Second,
the first term of the holographic entanglement entropy diverges and
is cutoff-dependent. The scaling of $a$ is given by the total
scalings of the $y$ and $w$ directions relative to that of the
$t$-direction: $(1+\frac{2}{3})/1=\frac{5}{3}$.

The second, finite term of the holographic entanglement entropy is
more interesting: it is cutoff-independent therefore can be compared
with the field theoretic computation of the entanglement entropy.
Its coefficient $\gamma_2$ gives a measure of the total degrees of
freedom. The scaling of $\ell_x$ is simply the total scalings of the
$y$ and $w$ directions relative to that of the $x$-direction:
$(1+\frac{2}{3})/1=\frac{5}{3}$. Since the scaling of the
$x$-direction is the same as that of the $t$-direction, the
exponents of $a$ and $\ell_x$ are the same.

\subsection{Entanglement Entropy for Subsystem along $w$-direction}
The next easiest subsystem we can consider is to divide along the
$w$-direction:
\begin{equation}
w \in [0,\ell_{w}<L_{w}].
\end{equation}
The three-dimensional minimal surface bordering on $\partial A$ is
given by the embedding function $z=z(w)$:
\begin{equation}
\mbox{Area}^{min}_{w}=\tilde{R}^3L_xL_{y}\int^{\ell_{w}}_{0}dw\frac{1}{z^{3}}\sqrt{z^{2/3}+z'^2}.
\end{equation}
Then we could follow the line of the previous subsystem along the
$x$-direction. The computation is straightforward but more
complicated so we present here instead a simpler route which
utilizes the result for the $x$-direction subsystem.

The coordinate transformation
\begin{equation}
z=\tilde{z}^{\frac{3}{2}}, \qquad
(t,x,y,w)=\frac{3}{2}(\tilde{t},\tilde{x},\tilde{y},\tilde{w}),
\end{equation}
results in the metric
\begin{equation}
ds^2=\left(\frac{3}{2}\tilde{R}\right)^2
\left(\frac{-d\tilde{t}^2+d\tilde{x}^2+d\tilde{y}^2}
{\tilde{z}^{3}}+\frac{d\tilde{z}^2}{\tilde{z}^{2}}+\frac{d\tilde{w}^2}{\tilde{z}^{2}}\right)+R^2ds^2_{X_5}.
\end{equation}
Thus we can simply use the result from the $x$-direction case, with
$d=\frac{8}{3}$ replaced by $d_{w}=4$. First, we write down the
dictionary between values in the original coordinates and the new
one.
\begin{enumerate}
\item In the new coordinates, the full boundary system is
\begin{equation}
\tilde{x}\in [0,\frac{2}{3}L_x], \qquad \tilde{y}\in
[0,\frac{2}{3}L_y],\qquad \tilde{w} \in [0,\frac{2}{3}L_{w}],
\end{equation}
while the subsystem $A$ is
\begin{equation}
\tilde{x}\in [0,\frac{2}{3}L_x], \qquad \tilde{y}\in
[0,\frac{2}{3}L_y],\qquad \tilde{w} \in
[0,\frac{2}{3}\ell_{w}<\frac{2}{3}L_{w}].
\end{equation}
\item The lattice spacing in the new coordinates is related to that in the old
coordinates by
\begin{equation}
\tilde{a}=a^{\frac{2}{3}}.
\end{equation}
\item The turning point of $z$ is
$\tilde{z}_{*}=\frac{2}{3}z_{*}$.
\end{enumerate}

Using the result from the $x$-direction case, we find the minimal
surface in the new coordinates to be
\begin{eqnarray}
Area^{min}&=&3\tilde{R}^3L_xL_{y}\frac{1}{d_{w}-1}\left[\frac{1}{\tilde{a}^{d_{w}-1}}
-\frac{1}{\tilde{z}_{*}^{d_{w}-1}}\frac{\sqrt{\pi}\Gamma(\frac{1}{2d_{w}}+\frac{1}{2})}{\Gamma(\frac{1}{2d_{w}})}\right].
\end{eqnarray}
After translated back into the original coordinates, it gives the
entanglement entropy of subsystem along the $w$-direction
\begin{eqnarray}
S_{EE-w}&=&\left(\frac{11}{12}\right)^3\frac{\pi^2}{\mbox{Vol}(X_5)}\cdot
N^2L_xL_{y}\frac{1}{D-1}\left[\frac{1}{a^{D-1}}-\left(\frac{3}{2}\right)^{d_{w}-1}\left(\frac{2}{\ell_{w}}\right)^{d_{w}-1}
\left(\frac{\sqrt{\pi}\Gamma(\frac{1}{2d_{w}}+\frac{1}{2})}{\Gamma(\frac{1}{2d_{w}})}\right)^{d_{w}}\right]\non\\
&=&N^2L_xL_{y}\left[\frac{\gamma_1'}{a^{2}}-\frac{\gamma_2'}{(\ell_{w})^{3}}\right],
\end{eqnarray}
where $D-1\equiv \frac{2(d_{w}-1)}{3}=2$. The negative finite part
has the same form as the result for the subsystem divided along the
$x$-direction with $d=\frac{8}{3}$ replaced by
$d_{w}=4$.\footnote{This might be understood as follows: when all
scalings are normalized with respect to the $w$-direction, then the
total scaling of the $(x,y,w)$-space is
\begin{equation}
\lambda_{x}+\lambda_y+\lambda_{w}=\frac{3}{2}+\frac{3}{2}+1=4.
\end{equation}
}

Now let's compare this result with the one for the subsystem along
the $x$-direction.  The essential features are the same. It is
proportional to the boundary area $L_xL_y$. There are one cut-off
dependent, divergent term and one cut-off independent, finite term.
The scaling of the cutoff $a$ is given by the total scalings of the
$x$ and $y$ directions relative to that of the $t$-direction:
$(1+1)/1=2$. The scaling of the $\ell_{w}$ is given by the total
scalings of the $x$ and $y$ directions relative to that of the
$w$-direction: $(1+1)/(2/3)=3$. Unlike the case for the subsystem
along the $x$-direction,  since the scaling of the $w$-direction is
different from that of the $t$-direction, the exponents of $a$ and
$\ell_{w}$ are different.

\section{Perturbative Analysis}
\label{sec:Pert}

In order to know the details of the holographic dual field theories,
a basic thing to do is to analyze the perturbative spectra around
their supergravity solutions. This offers us the information on
scale dimensions \cite{ADSGKP,ADSWitten}. For example, scalar
perturbations are dual to scalar operators ${\cal O}_i$ in the dual
field theory. These perturbative modes in supergravity are described
by Klein-Gordon equations with various masses in the curved
spacetime. Since we have the nontrivial dilaton in our D3-D7 scaling
solutions (\ref{scalingmetric}), it is not clear a priori whether
the Klein-Gordon equation should be obtained from the string frame
metric or the Einstein frame metric. Actually our scaling property
(\ref{scalingt}) of the gravity solutions is only available in the
Einstein frame. Also the study of the perturbative spectrum is
necessary to determine the stability of the background. Motivated by
these, below we will examine the perturbations around our scaling
backgrounds (\ref{scalingmetric}).

\subsection{Description of Perturbations}

Let us analyze the perturbations of bosonic fields around the D3-D7
scaling solutions defined by (\ref{scalingmetric}),
(\ref{pdilaton}), (\ref{fivef}) and (\ref{onef}) in the Einstein
frame . We will closely follow the analysis of $AdS_5\times S^5$ in
\cite{KRN,DFGHM}. We will denote the total ten-dimensional
coordinates by $M,N,..=0,1,2,\ddd,9$. The five-dimensional
Lorentzian spacetime (called $M_5$) is described by the coordinate
$\mu,\nu,\ddd =0,1,2,3,4$ and the five-dimensional compact manifold
$X_5$ by $\ap,\beta,\ddd =5,6,7,8,9$.

As is clear from the IIB supergravity action in our background, the
3-form fluxes $F_3$ and $H_3$ are decoupled from the other fields
(i.e. the metric, the dilaton, $F_5$, and $F_1$) thus we can
concentrate on the latter ones. Then the Lagrangian in the Einstein
frame is written as
follows \be {\cal
L}=\s{-g}\left(R-\f{1}{2}e^{2\phi}\de_M\chi\de^M\chi
-\f{1}{2}\de_M\phi\de^M\phi-\f{1}{4\cdot
5!}F_{MNPQR}F^{MNPQR}\right). \label{actionein} \ee

To make analysis more tractable, let us assume $X_5=S^5$ here. Then
we can define the scalar, vector, traceless symmetric, and
antisymmetric spherical harmonics on $S^5$ by $Y^I$, $Y^I_\ap$,
$Y^I_{(\ap\beta)}$, and $Y^I_{[\ap\beta]}$, respectively \cite{KRN}.
Using these spherical harmonics, the metric perturbations $\delta
g_{MN}=h_{MN}$ can be decomposed as follows \cite{KRN} \ba
h_{(\mu\nu)}=h_{(\mu\nu)}^I Y^I, \ \ \ \ h^\mu_\mu=h^IY^I,\ \ \ \
h_{\mu\ap}=B^I_\mu Y^I_\ap,\ \ \ \
h_{(\ap\beta)}=\phi^IY^I_{(\ap\beta)},\ \ \ \ h^\ap_\ap=\pi^IY^I,
\label{pertmetric} \ea where $(\ap\beta)$ denotes the traceless
symmetric part. We also denoted all indices of the spherical
harmonics simply by $I$. We fix the gauge by requiring \be
\nabla^\ap h_{(\ap\beta)}=\nabla^\ap h_{\mu\ap}=0. \label{gauge} \ee

The perturbations of the dilaton and axion are defined as follows:
\be
\delta \phi=\vp^IY^I,\ \ \ \ \delta \chi=\eta^IY^I. \label{pertdilaton}
\ee

Finally, the perturbation of the 5-form flux $F_5=dC_4$ can be
express as follows  \cite{KRN}: \ba &&
C_{\mu_1\mu_2\mu_3\mu_4}=b^I_{\mu_1\mu_2\mu_3\mu_4}Y^I,\ \ \ \
 C_{\mu_1\mu_2\mu_3\ap}=b^I_{\mu_1\mu_2\mu_3}Y^I_{\ap},\ \ \ \
 C_{\mu_1\mu_2\ap_1\ap_2}=b^I_{\mu_1\mu_2}Y^I_{[\ap_1\ap_2]},\no
&& C_{\mu\ap_1\ap_2\ap_3}=b^I_\mu \ep_{\ap_1\ap_2\ap_3}^{\ \ \ \ \ \ \ \beta_1\beta_2}
\nabla_{\beta_1}Y^I_{\beta_2},\ \ \ \
 C_{\ap_1\ap_2\ap_3\ap_4}=b^I\ep_{\ap_1\ap_2\ap_3\ap_4}^{\ \ \ \ \ \ \ \ \ \  \beta}\nabla_\beta Y^I.
 \label{pertfive}
\ea
They satisfy the gauge fixing condition $\nabla^\ap C_{\ap\ddd}=0$, and the self-duality of
$F_5$ allows us to eliminate $b^I_{\mu_1\mu_2\mu_3\mu_4}$ and $b^I_{\mu_1\mu_2\mu_3}$.

Next we substitute (\ref{pertmetric}), (\ref{pertdilaton}), and
(\ref{pertfive}) into equations of motion of (\ref{actionein}) and
derive the perturbative differential equations. We omit the details
of analysis here and put them in Appendix \ref{Aperturbation} as
many parts of the calculations are essentially the same as those in
\cite{KRN}.

In the end, we find that the following modes \ba &&\mbox{Scalar
modes}: \phi^I,\  (h^I,\pi^I,b^I),\no &&\mbox{Vector modes}:
(B^I_\mu,b^I_\mu),\no &&\mbox{Tensor modes}: b^I_{\mu\nu},
\label{simplemodes} \ea satisfy free massive field equations which
are precisely the same expressions as those in the $AdS_5\times S^5$
background.\footnote{ In other words, for these modes the
differences from $AdS_5\times S^5$ only come from the background metric which
is employed to write down the free field equations.} In the above,
the fields in the same parenthesis mix with each other. Therefore,
these perturbations (\ref{simplemodes}) obey free field equations of
motion constructed from the Einstein frame metric
(\ref{scalingmetric}) instead of the string frame metric.

As an example, let us concentrate on the scalar mode $\phi^I$. Its
equation of motion is written as \be
\left(\Box_x+\Box_y-\f{2}{R^2}\right)\phi^IY^I_{(\ap\beta)}=0, \ee
where $\Box_x$ and $\Box_y$ are the Laplacians of the Lorentzian
part $M_5$ and the sphere part $S^5$, respectively. Using the
eigenvalues of $Y^{I}_{(\ap\beta)}$, we eventually obtain\footnote{
Please distinguish the total angular momentum $k$ of the spherical harmonics of $S^5$
from the number $k$ of D7-branes.}
\be
\left(\Box_x-\f{k(k+4)}{R^2}\right)\phi^I=0,\ \ \ (k=2,3,4,\ddd).
\label{exphi} \ee

On the other hands, the other modes $\vp^I$, $\eta^I$, and
$h_{\mu\nu}$ mix with each other and obey equations of motion more
complicated than those in the $AdS_5\times S^5$ case (see Appendix
\ref{Aperturbation}).

\subsection{Scaling Dimensions and Stability}

We have observed that a large class of supergravity modes
(\ref{simplemodes}), though not all of them,  satisfy the
conventional free field equations with various masses via the
Kaluza-Klein compactification on $S^5$. The scalar modes in
(\ref{simplemodes}) satisfy the equations of motion of the form
(we denote such a scalar field by $\Phi$ here) \be
\left(\Box_x-m^2\right)\Phi=0,\label{Phieom} \ee where the Laplacian
$\Box_x=g^{\mu\nu}\nabla_\mu\nabla_\nu$ is constructed from the
Einstein frame metric (\ref{scalingmetric}).

Now let us consider a scalar field $\Phi$ on a slightly generalized
scaling background \be
ds^2=\ti{R}^2\left(\f{dz^2}{z^2}+\f{-dt^2+dx^2+dy^2}{z^2}+\f{dw^2}{z^{2\nu}}\right),\label{nuba}
\ee where the scaling exponent $\nu$ is related to the scaling
exponent $z$ in (\ref{aniso}) by $\nu=\f{1}{z}$. Especially, our scaling
background (\ref{scalingmetric}) corresponds to $\nu=\f{2}{3}$. The
equation of motion (\ref{Phieom}) is written as follows: \be
-\Phi''+\f{\nu+2}{z}\Phi'+\left(\f{m^2\ti{R}^2}{z^2}+p^2-\omega^2+p_w^2~z^{2(\nu-1)}\right)\Phi=0.
\ee Here $\omega$, $p$, and $p_w$ are the frequency,
the momenta in $(x,y)$- and $w$-direction, respectively.

After redefining the wave function by
$\Phi(z)=z^{\f{\nu+2}{2}}\Psi(z)$, we obtain the Schrodinger-like
equation \be -\Psi''+V(z)\Psi=\omega^2\Psi, \ee where \be
V(z)=\f{m^2\ti{R}^2+\f{(3+\nu)^2-1}{4}}{z^2}+p^2+p_w^2~z^{2(\nu-1)}.
\label{potentiale} \ee When $z$ is small, the third term in
(\ref{potentiale}) is small compared to the first term, assuming
$\nu>0$. Thus, as in the AdS/CFT case (i.e. $\nu=1$), we can
expect\footnote{We are very grateful to Andreas Karch for illuminating
explanations on the stability analysis in Poincar\'{e} AdS spaces.} that the
stability against the (normalizable) perturbations is the same as
that of Schrodinger problem with the potential
$V(z)=\f{\lambda-\f{1}{4}}{z^2}$. It is well-known that the latter
system is stable iff $\lambda\geq 0$.

In this way, we speculate that in the background (\ref{nuba}), the stability condition requires
\be
m^2\ti{R}^2\geq -\f{(3+\nu)^2}{4}.  \label{stabilityc}
\ee
Notice that if we set $\nu=1$ in (\ref{stabilityc}), we obtain $m^2\ti{R}^2\geq -4$, which is the
well-known Breitenlohner-Freedman (BF) bound of $AdS_5$.

This condition can equally be implied from the behavior of the
scalar field near the boundary $z\to 0$ \be \phi(z)\sim
Az^{\Delta_+}+Bz^{\Delta_-}+\ddd, \ee where \be
\Delta_{\pm}=\f{\nu+3}{2}\pm \s{\f{(\nu+3)^2}{4}+m^2\ti{R}^2}. \ee
The quantity $\Delta_{\pm}$ is holographically interpreted as the
scaling dimension in the dual anisotropic scale invariant theory.
Notice that the condition (\ref{stabilityc}) requires that the
scaling dimension should be real-valued.

Now let us go back to our D3-D7 scaling solutions
(\ref{scalingmetric}). In this case we obtain the stability
condition by setting $\nu=2/3$ and $R^2=\f{11}{12}\ti{R}^2$ as
follows \be m^2R^2\geq -\f{11}{3}. \label{boundsc} \ee We can apply
this condition to the scalar modes in (\ref{simplemodes}). As is
clear from (\ref{exphi}), all the scalar modes $\phi^I$ satisfy this
condition. However, we find that one of the infinitely many mixed
modes of $(h^I,\pi^I,b^I)$ actually has the largest tachyonic mass
$m^2R^2=-4$, which saturates the BF bound of $AdS_5$. This occurs
only for the second spherical harmonics $k=2$. Even though this
tachyonic mode is stable in the $AdS_5$ spacetime, it seems to
become an unstable mode in our D3-D7 scaling background as
(\ref{boundsc}) is violated.\footnote{A perturbative instability has
also been noticed in \cite{HaYo} for type IIB backgrounds dual to
the non-relativistic CFT when the 3-form fluxes are vanishing.}

Nevertheless, we can replace $S^5$ with an arbitrary Einstein
manifolds $X_5$ with the same Ricci curvature, keeping the same
scaling solution (\ref{scalingmetric}). Define the eigenvalues
$\Lambda$ of Laplacian of a scalar function $Y$ such that
$-R^2\Box_y Y=\Lambda Y$. In this case, if \be
\f{\Lambda}{16}+1-\s{\f{\Lambda}{4}+1}\geq -\f{11}{48}, \ee is always
satisfied, then the above lowest mass mode becomes stable. In other
words, if there is no eigenvalue between \be \f{37-8\s{3}}{3}(\simeq
7.71)< \Lambda<\f{37+8\s{3}}{3}(\simeq 16.95), \ee then the
background can be perturbatively stable. Notice that $\Lambda=12$
saturates the BF bound and it is the unstable mode when $X_5=S^5$.
It is intriguing whether there exists such a stable
(Sasaki-)Einstein manifold.

\section{D4-D6 Scaling Solutions}
\label{sec:dfds}

In type IIA string, the closest analogue to the previous D3-D7
system is the following D4-D6 system:
\begin{equation}\label{D4-D6}
\begin{array}{r|cccc|cc|cccccl}
\,\, \mbox{$\mathcal{M}_4\times T^2\times X_4$:}\,\,\, & t     & x
& y & r  & w_1 & w_2   & s_1   & s_2   & s_3   & s_4 & \, \nonumber\\
\hline N \,\, \mbox{D4:}\,\,\,& \x &   \x  &  \x &   & \x & \x  &  &
& &
&\,\nonumber\\
k\,\, \mbox{D6:}\,\,\,& \x &   \x  &  \x &   &  &   & \x & \x &
\x&\x &\,\nonumber
\end{array}
\end{equation}
Here $(w_1,w_2)$ span a two-manifold that supports the D6 flux (we
will choose it to be $T^2$ for simplicity) and $(s_1,s_2,s_3,s_4)$
span a four-dimensional Einstein manifold $X_4$ with the same Ricci
curvature as that of $S^4$. However, as we will see, this system
does not support a scaling-invariant solutions. In this section, we
will present a gravity solution that is closest to a scaling
solution: under $(t,x,y,r,w_1,w_2) \rightarrow (\lambda t, \lambda
x, \lambda y, \frac{r}{\lambda}, \lambda^{\frac{2}{3}} w_1,
\lambda^{\frac{2}{3}} w_2)$, the line element $ds^2\rightarrow
\lambda^{-\frac{1}{3}}ds^2$. We will also give its black brane
generalization.

The fluxes given by these $N$ D4-branes and $k$ D6-branes are:
\begin{equation}
F_{2}=\frac{(2\pi)k}{L^2}dw_1\wedge dw_2,\qquad \quad F_{4}=(2\pi)^3
N\frac{1}{\mbox{Vol}(X_4)}\Omega_{X_4},\qquad \quad B_{2}=0,
\end{equation}
where $X_4$ is a unit-radius Einstein 4-manifold (whose Ricci tensor
satisfies $R_{ij}=3g_{ij}$) and $\Omega_{X_4}$ is its volume-form; $L$
is the periodicity of $w_i$. This flux profile satisfies the flux
equations of motion. The corresponding (string frame) metric ansatz
is:
\begin{eqnarray}
ds^2_s&=&e^{2B(r)}(-dt^2+dx^2+dy^2)+e^{-2A(r)}dr^2+e^{2T(r)}(dw_1^2+dw_2^2)+e^{2Z(r)}ds^2_{X_4},\nonumber\\
\end{eqnarray}
with a possibly non-constant dilaton $\phi(r)$.

For a scaling solution, $\{A,B,T,Z,\phi\}$ are
\begin{eqnarray}
&&A(r)=a_1 \log{r}+a_0,\qquad B(r)=b_1 \log{r}+b_0, \qquad T=t_1
\log{r}+t_0,\nonumber\\ && Z=z_1 \log{r}+z_0, \qquad \phi(r)=\eta_s
\log{r}+\phi_0.
\end{eqnarray}
There are one equation of motion from the dilaton and five from the
gravity part. For the scaling ansatz, they all reduce to algebraic
equations and the solution is easily found to be:
\begin{eqnarray}
A(r)&=&\left(1-\frac{1}{3}\eta_s\right)\log(r)-\frac{1}{3}\log{\left(\frac{8\pi^3}{3\mbox{Vol}(X_4)} e^{\phi_0}N\right)}-\frac{1}{2}\log{\left(\frac{17}{ 18}\right)}-\log{\eta_s},\non\\
B(r)&=&\frac{5}{6}\eta_s\log{r}+b_0,\non\\
T(r)&=&\frac{2}{3}\eta_s\log{r}+\frac{1}{6}\log{\left(\frac{8\pi^3}{3\mbox{Vol}(X_4)}e^{\phi_0}N\right)}
+\frac{1}{2}\log{\left(  \frac{2\pi}{L^2}e^{\phi_0}k\right)},\\
Z(r)&=&\frac{1}{3}\eta_s\log(r)+\frac{1}{3}\log{\left(\frac{8\pi^3}{3\mbox{Vol}(X_4)}  e^{\phi_0}N\right)},\non\\
\phi(r)&=&\eta_s\log{r}+\phi_0.\non
\end{eqnarray}
$(\eta_s,b_0,\phi_0)$ are three gauge parameters. $\eta_s$
corresponds to the gauge freedom of $r\rightarrow r^{\alpha}$, $b_0$
corresponds to that of rescaling the $(t,x,y)$ directions, and
$\phi_0$ gives the string coupling at $r=1$ thus corresponds to
rescalings of the $r$ and $T^2$ directions.

Without loss of generality, we choose
\begin{equation}
\eta_s=2, \qquad
b_0=\frac{1}{3}\log{\left(\frac{8\pi^3}{3\mbox{Vol}(X_4)}
e^{\phi_0}N\right)} +\frac{1}{2}\log{\frac{34}{9}},
\qquad\phi_0=\frac{1}{2}\log{\left(\frac{L^6 N}{3\mbox{Vol}(X_4)
k^3}\right)}+\frac{3}{2}\log{\frac{34}{9}}.
\end{equation}
The solution in the string frame is
\begin{eqnarray}
ds_s^2&=&\tilde{R}_s^2\left[r^{\frac{10}{3}}(-dt^2+dx^2+dy^2)
+\frac{dr^2}{r^{\frac{2}{3}}} +r^{\frac{8}{3}}(dw_1^2+dw_2^2)\right]
+R_s^2r^{\frac{4}{3}}ds^2_{X_4},
\end{eqnarray}
with
$R_s^2=\frac{9}{34}\tilde{R}_s^2=\frac{8\pi^2}{3\mbox{Vol}(X_4)}\cdot\frac{17}{9}\cdot
N\frac{L^2}{k}$. In the Einstein frame, it is
\begin{equation}
ds_E^2=\tilde{R}^2\left[r^{\frac{7}{3}}
(-dt^2+dx^2+dy^2)+\frac{dr^2}{r^{\frac{5}{3}}}
+r^{\frac{5}{3}}(dw_1^2+dw_2^2)\right] +R^2r^{\frac{1}{3}}ds^2_{X_4},
\end{equation}
with $R^2=\frac{9}{34}\tilde{R}^2=(
\frac{8\pi^2}{3\mbox{Vol}(X_4)})^{\frac{3}{4}}
(\frac{17\pi^2}{9}\cdot N^3\frac{L^2}{k})^{1/4}$.

This background is no longer scaling invariant. Under the scaling
transformation
\begin{equation}
(t,x,y,r,w_1,w_2) \rightarrow (\lambda t, \lambda x, \lambda y,
\frac{r}{\lambda}, \lambda^{\frac{2}{3}} w_1, \lambda^{\frac{2}{3}}
w_2),
\end{equation}
the metric scales as
\begin{equation}
ds^2 \rightarrow \lambda^{-\frac{1}{3}}ds^2,
\end{equation}
instead of staying invariant. This is not surprising since the
D4-brane theory is not conformal in the first place.

Generalizing to finite temperature, the corresponding black brane
solution (in the Einstein frame) is
\begin{equation}
ds_E^2=\tilde{R}^2\left[r^{\frac{7}{3}}(-F(r)dt^2+dx^2+dy^2)
+\frac{dr^2}{F(r)r^{\frac{5}{3}}}+r^{\frac{5}{3}}
(dw_1^2+dw_2^2)\right]+R^2r^{\frac{1}{3}}ds^2_{X_4},
\end{equation}
with
\begin{equation}
F(r)=1-\frac{\mu}{r^{\frac{17}{3}}},
\end{equation}
where $\mu$ is the mass parameter of the black hole. Its Hawking
temperature is
\begin{equation}
T_H=\frac{17 }{12\pi}\mu^{\frac{3}{17}}.
\end{equation}
The Bekenstein-Hawing entropy is
\begin{equation}
S=\gamma\cdot \left(\frac{8\pi^2}{3\mbox{Vol}(X_4)}\right)^2 \cdot
T_H^{\frac{14}{3}}\cdot N^3 \cdot V_2 \cdot \frac{L^4}{k},
\end{equation}
with $\gamma=2^{\frac{28}{3}}\cdot 3^{-\frac{7}{3}}\cdot
17^{-\frac{5}{3}}\cdot \pi^{\frac{8}{3}}$ and $V_2$ is the area in
the $(x,y)$ directions.

\section{Conclusions and Discussions}
\label{sec:conc}

In this paper, we presented a class of gravity duals of
Lifshitz-like fixed points in type IIB supergravity. They represent
backgrounds with intersecting D3 and D7-branes and their Einstein
frame metrics (\ref{scalingmetric}) enjoy anisotropic scale
invariance. We also extended them to black brane solutions dual to
finite temperature theories. Moreover, we showed the existence of
solutions which interpolate between our anisotropic solutions in the
IR and the familiar $AdS_5\times X_5$ solutions in the UV. Then the
holography asserts that our Lifshitz-like fixed points can be
obtained from various four-dimensional CFTs including ${\cal N}=4$
super Yang-Mills via RG flows. These flows are triggered by the
relevant and anisotropic perturbation which makes the $\theta$-angle
(the coefficient in front of the topological Yang-Mills coupling
$F\we F$) linearly dependent on one of the space-like coordinates
i.e. $\theta\propto w$. When $w$ is compactified, the perturbation
induces the Chern-Simons coupling $\int A\we F+\f{2}{3}A^3$. This
theory itself seems an intriguing model worth pursuing in a future
work, as the equation of motion becomes local in spite of the
violation of the Lorentz invariance.

Employing our supergravity solutions we studied the thermal entropy
and the entanglement entropy to measure the degrees of freedom of
the holographic dual theories. We found characteristic scaling
properties in both quantities. We also holographically computed the
shear and bulk viscosities. A more general analysis of hydrodynamics
with the momentum in the $w$-direction taken into account was left
as an interesting future problem.

Moreover, we performed a perturbative analysis around our solutions
and found that a large class of scalar modes obey the Klein-Gordon
equation in the curved spacetime which has the expected scaling
property. Also we found an unstable scalar mode when the compact
manifold $X_5$ is $S^5$. Since this unstable mode occurs only for a
`d-wave' spherical harmonics, $S^5$ might decay into a less
symmetric Einstein manifold and be stabilized. Thus we have reason
to hope that there exists a class of (Sasaki-)Einstein manifolds
with which our scaling solutions become stable. Even the background
with $X_5=S^5$ is still useful at least in capturing qualitative
properties of gravity duals of Lifshitz-like fixed points, with
unstable modes simply neglected. The construction of manifestly
stable and non-dilatonic gravity duals of Lifshitz-like fixed points
still remains as a very interesting future problem. The three-form
fluxes which we assume to be vanishing in our solution might play an
important role.

It is also intriguing to generalize our solutions to other values of
$p$ and $d$ in (\ref{AdSani}). For example, the simplest case $p=0$
deserves particular attention. It can be formally obtained from our
solution (\ref{scalingmetric}) by the double Wick rotation $t\to iw$
and $w\to it$. However, the axion field $\chi$ and its flux become
imaginary-valued therefore the solution is not physical in the
ordinary type IIB supergravity. A slightly better example which
realizes the case $p=0$ is a background based on D3-D5 systems,
where D5-branes are regarded as the baryons \cite{Witten}. We cannot
get any consistent solution if we restrict to the ordinary type IIB
supergravity because the tadpole for the $H$-flux is generated by
F-stings which attach to D5-branes and stretch into the boundary
\cite{Witten}. To construct a solution in this background we need to
add the F-string action as an extra term to the type IIB
supergravity. Under this slightly unusual assumption, we can indeed
find the following black brane solution in the Einstein frame
\cite{ANT} \ba &&
ds^2_E=-A(\rho)\rho^{14}dt^2+\rho^2(dx^2+dy^2+dz^2)+\ti{L}^2\f{d\rho^2}{\rho^2A(\rho)}
+L^2d\Omega^2,\no && e^{\phi(\rho)}=e^{\phi_0}\rho^6,\ \ \ \ \
A(\rho)=1-\f{\mu}{\rho^{10}}, \ \ \ \ \ \ti{L}^2=10L^2,
\label{baryonm} \ea with constant 3-form fluxes $H_3,F_3$ and the RR
5-form $F_5$. A derivation of this solution is briefly reviewed in
Appendix \ref{sec:baryon}. At zero temperature, this corresponds to
the metric (\ref{AdSani}) with $p=0$, $d=3$ and $z=7$.

Finally, it is also intriguing to apply our backgrounds to realistic
condensed matter systems. Our D3-D7 model was originally introduced
to model the holographic dual of fractional quantum Hall effects in
string theory \cite{FLRT} (refer to \cite{Kraus,KrausT,HLT} for
other holographic realizations of quantum Hall effects). Therefore
one of the future problems is to calculate physical quantities such
as finite temperature corrections to the Hall and longitudinal
conductivities in this theory. The anisotropic critical points we
found in this paper may also be useful to analyze the systems like liquid
crystals and some anisotropic spin systems.

\vskip8mm

\noindent {\bf Acknowledgments}

We thank very much S. Hellerman, Y. Hikida, A. Karch, E. Kiritsis,  S.
Mukohyama, T. Nishioka, S. Ryu, W. Song, S. Sugimoto and K. Yoshida
for useful discussions. We are also benefited from
conversations with D. Gao, D. Orlando, S. Reffert and M.
Yamazaki. TA is supported by the Japan Society for the Promotion of
Science (JSPS) and by the Grant-in-Aid for the Global COE program
"The Next Generation of Physics, Spun from Universality and
Emergence" from the MEXT. WL and TT are supported by World Premier
International Research Center Initiative (WPI Initiative), MEXT,
Japan.
 The work of TT is also supported by JSPS Grant-in-Aid for
Scientific Research No.20740132 and by JSPS
Grant-in-Aid for Creative Scientific Research No. 19GS0219.

\vskip2mm


\newpage

\appendix
\settocdepth{section}

\section{Perturbative Analysis}
\label{Aperturbation}

In this appendix, we consider perturbative fluctuations around the
D3-D7 background \eqref{fivef}, \eqref{onef}, \eqref{pdilaton}, and
\eqref{scalingmetric} in the Einstein frame with $X_5=S^5$, and analyze the
stability of it by using the linearized equations of motion. Our
analysis below mostly follows \cite{KRN, DFGHM}. Here we notice that
RR 3-form flux $F_3$ and NSNS 3-form flux $H_3$ vanish on this
background, and their fluctuations do not mix with that of the other
fields in the linear order. Thus we start with the action
\eqref{actionein}. The fluctuation of the metric, dilaton, axion,
and RR 5-form flux $dC_4=F_5$ along with their decompositions in
terms of spherical harmonics on $S^5$ are summarized in
\eqref{pertmetric}, \eqref{pertdilaton}, and \eqref{pertfive}. Here
we consider the fluctuations satisfying the gauge fixing conditions
$\nabla^{\alpha}h_{(\alpha\beta)}=\nabla^{\alpha}h_{\mu\alpha}=0$
for the metric and
$\nabla^{\alpha}C_{\alpha IJKL}=0$ for the RR
5-form flux $F_5=dC_4$. For simplicity, we denote the background
metric, dilaton, axion, and 5-form flux by $g_{MN}$, $\phi$, $\chi$,
and $F_5$, respectively. In this section we define
$\tilde{\alpha}=\alpha/R^5=4/R$.

\subsection{Some Conventions}

Before writing down linearized equations of motion, we summarize
some of our conventions.

We normalized the $\ep$ tensor on the five-dimensional Lorentzian
part $M_5$ and the $S^5$ part in \eqref{scalingmetric} by \ba &&
\ep_{01234}=\s{-g_{M_5}},\ \ \ \
\ep^{01234}=-\f{1}{\s{-g_{M_5}}},\no && \ep_{56789}=\s{g_{S^5}},\ \
\ \ \ep^{56789}=\f{1}{\s{g_{S^5}}}. \ea In this convention, \be
\ep_{\mu_1\mu_2\mu_3\mu_4\mu_5}\ep^{\mu_1\mu_2\mu_3\mu_4\mu_5}=-5!\
,\ \ \ \ \ \ \ \
\ep_{\ap_1\ap_2\ap_3\ap_4\ap_5}\ep^{\ap_1\ap_2\ap_3\ap_4\ap_5}=5!\
.\ \ \ \ee We also define the ten-dimensional $\ep$ tensor by \be
\ep_{0123456789}=\s{-g_{10}}=\ep_{01234}\cdot\ep_{56789}. \ee It is
also useful to  define the Laplacian for $M_5$ and $S^5$ by \be
\Box_x=g^{\mu\nu}\nabla_\mu\nabla_\nu,\ \ \ \ \
\Box_y=g^{\ap\beta}\nabla_\ap\nabla_\beta. \ee

\subsection{Spherical Harmonics on $S^5$}

In this appendix, we decompose the linearized equations of motion
for the fluctuations by using the spherical harmonics on $S^5$. Thus
we define $Y^I$, $Y^I_\ap$, $Y^I_{(\ap\beta)}$, and
$Y^I_{[\ap\beta]}$, which represent scalar, vector,traceless
symmetric, and antisymmetric spherical harmonics, respectively. They
satisfy the transverse conditions \be \nabla^\ap Y^I_\ap=\nabla^\ap
Y^I_{(\ap\beta)}=\nabla^\ap Y^I_{[\ap\beta]}=0. \ee

For the vector spherical harmonics, we define the Hodge-de Rham
operator $\Delta_y$ by $\Delta_y Y^I_\beta=\Box_y
Y^I_\beta-R^\ap_\beta Y^I_\ap$. We can define the Hodge-de Rham
operators for the other spherical harmonics in the similar manner
and the eigenvalues of $\Delta_y$ on the $S^5$ with the radius $R$
are given by\footnote{
Please distinguish the total angular momentum $k$ of the spherical harmonics of $S^5$
from the number $k$ of D7-branes.} \cite{KRN} \ba && \Delta_y Y^I=\Box_y
Y^I=-\f{k(k+4)}{R^2}Y^I,  \ \ \ (k=0,1,2,\ddd)
\label{harm_scalar}\\
&& \Delta_y Y^I_\ap=\left(\Box_y-\f{4}{R^2}\right) Y^I_{\ap}
=-\f{(k+1)(k+3)}{R^2}Y^I_\ap,  \ \ \ (k=1,2,\ddd)
\label{harm_vector}\\
&& \Delta_y Y^I_{(\ap\beta)}=\left(\Box_y-\f{10}{R^2}\right)
Y^I_{(\ap\beta)} =-\f{k^2+4k+8}{R^2}Y^I_{(\ap\beta)},  \ \ \
(k=2,3,4,\ddd)
\label{harm_sym}\\
&& \Delta_y Y^I_{[\ap\beta]}=\left(\Box_y-\f{6}{R^2}\right)
Y^I_{[\ap\beta]} =-\f{(k+2)^2}{R^2}Y^I, \ \ \
(k=1,2,\ddd).\label{harm_antisym} \ea

\subsection{Five-form Flux Equation}
\label{subsec:linear_five}

Let us first consider the perturbation for the RR 5-form flux $F_5$.
It satisfies the self-duality constraint \be
F_{PQRST}=\f{1}{5!}\ep_{PQRST}^{\ \ \ \ \ \ \ \
ABCDE}F_{ABCDE}.\label{selfdual} \ee from which the equation of
motion $d*F_5=0$ follows automatically. By denoting the fluctuation
of the 5-form around the background as $\delta F_5=f_5$,
(\ref{selfdual}) in the linear order is written as \ba &&
f_{PQRST}=\f{1}{5!}\ep_{PQRST}^{\ \ \ \ \ \ \ \
ABCDE}f_{ABCDE}+\f{1}{2\cdot 5!}h^{\ M}_M \ep_{PQRST}^{\ \ \ \ \ \ \ \
ABCDE}F_{ABCDE}\no &&\ \ \ \  \ \ \ \ \ \ \ \  \ \ \ \ \ -\f{1}{4!}
\ep_{PQRSTA_1}^{\ \ \ \ \ \ \ \ \ \  BCDE}F_{A_2BCDE}h^{A_1A_2}. \ea
By substituting \eqref{pertmetric} and \eqref{pertfive} into this
equation and then decomposing it by the spherical harmonics on
$S^5$, we obtain five equations: \ba && \left[5\nabla_{\mu_1}
b^I_{\mu_2\mu_3\mu_4\mu_5}-\ep_{\mu_1\mu_2\mu_3\mu_4\mu_5}
\left(\f{\ti{\ap}}{2}h^I-\f{\ti{\ap}}{2}\pi^I+b^I\Box_y\right)\right]Y^I=0, \label{fluxone}\\
&& \left[4\nabla_{\mu_1}
b^I_{\mu_2\mu_3\mu_4}+\ep_{\mu_1\mu_2\mu_3\mu_4}^{\ \ \  \ \ \ \ \ \
\nu}
(b^I_\nu\Box_y-\ti{\ap}B^I_\nu)\right]Y^I_\ap=0,\label{fluxtwo}\\
&&\left[b^I_{\mu_1\mu_2\mu_3\mu_4}+\ep_{\mu_1\mu_2\mu_3\mu_4}^{\ \ \
\ \ \ \ \ \ \nu}
\nabla_\nu b^I\right]\nabla_\ap Y^I=0,\label{fluxthree}\\
&& \nabla_{\mu_1}
b^I_{\mu_2\mu_3}Y^{I}_{[\ap\beta]}-\f{1}{12}\ep_{\mu_1\mu_2\mu_3}^{\
\ \ \ \ \ \ \nu_1\nu_2}
b^I_{\nu_1\nu_2}\ep_{\ap\beta}^{\ \ \gamma_1\gamma_2\gamma_3}\nabla_{\gamma_1}Y^I_{[\gamma_2\gamma_3]}=0,\label{fluxfour}\\
&& \left[b^I_{\mu_1\mu_2\mu_3}+\ep_{\mu_1\mu_2\mu_3}^{\ \ \ \ \ \
\nu_1\nu_2} \nabla_{\nu_1}b^I_{\nu_2}\right]\nabla_{[\ap}
Y^I_{\beta]}=0.\label{fluxfive} \ea Here
$\tilde{\alpha}=\alpha/R^5=4/R$. Now, we can simply solve
(\ref{fluxthree}) and (\ref{fluxfive}) algebraically, assuming
$k\geq 1$. Then, we obtain the following three equations from
\eqref{fluxone} and \eqref{fluxthree}, \eqref{fluxtwo} and
\eqref{fluxfive}, and \eqref{fluxfour} respectively: \ba
&& \left[(\Box_x+\Box_y)b^I+\f{\ti{\ap}}{2}h^I-\f{\ti{\ap}}{2}\pi^I\right]Y^I=0,\label{ffluxone}\\
&& \left[\Box_xb^I_\mu-\nabla^\nu\nabla_\mu b^I_\nu+\Delta_y
b^I_\mu-\ti{\ap}B^I_\mu\right]Y^I_\ap=0,
\label{ffluxtwo}\\
&& \left[3\nabla_{\mu_1} b^I_{\mu_2\mu_3}\mp
\f{i}{2}\ep_{\mu_1\mu_2\mu_3}^{\ \ \ \ \ \ \ \nu_1\nu_2}
b^I_{\nu_1\nu_2}\s{-\Delta_y}\right]Y^{\pm}_{[\ap\beta]}=0.\label{ffluxthree}
\ea

\subsection{Einstein Equations}
\label{subsec:linear_einstein}

Let us next consider the perturbation for the Einstein equation. The
Einstein equation itself can be obtained from the action
(\ref{actionein}) as \ba &&
R_{MN}-\f{1}{2}g_{MN}\left(R-\f{1}{2}e^{2\phi}\de_P\chi\de^P\chi-\f{1}{2}\de_P\phi\de^P\phi-\f{1}{4\cdot
5!} F_{PQRST}F^{PQRST}\right)\no && \ \ \ \ \
-\f{1}{2}e^{2\phi}\de_M\chi\de_N\chi-\f{1}{2}\de_M\phi\de_N\phi
-\f{1}{4\cdot 4!}F_{MPQRS}F_{N}^{\ \ PQRS}=0. \label{einone} \ea By
using \be
R-\f{1}{2}e^{2\phi}\de_P\chi\de^P\chi-\f{1}{2}\de_P\phi\de^P\phi=0.
\ee and \be F_{PQRST}F^{PQRST}=0, \ee derived from the trace part of
\eqref{einone} and the self-duality condition \eqref{selfdual}
respectively, we can reduce the Einstein equation (\ref{einone}) to
a simpler form: \be
 R_{MN}-\f{1}{2}e^{2\phi}\de_M\chi\de_N\chi-\f{1}{2}\de_M\phi\de_N\phi
-\f{1}{4\cdot 4!}F_{MPQRS}F_{N}^{\ \ PQRS}=0. \label{ein} \ee By
linearly perturbing \eqref{ein} and decomposing it in terms of the
spherical harmonics on $S^5$, we obtain some equations for the
fluctuations around the background. We summarize the resulting
equations below.

\subsubsection{$\ap\beta$ Components}

From $(\alpha,\beta)$-components of \eqref{ein}, we obtain the
following four equations: \ba
&&\left[\Box_x+\Box_y-\f{2}{R^2}\right]\phi^IY^I_{(\ap\beta)}=0,\label{EinABone}\\
&&\nabla^\mu B^I_\mu\nabla_{(\ap}Y^I_{\beta)}=0,\label{EinABtwo}\\
&&\left[h^I+\f{3}{5}\pi^I\right]\nabla_{(\ap}\nabla_{\beta)}Y^I=0,\label{EinABthree}\\
&& \left[\f{1}{10}\Box_x\pi^I +\f{4}{25}\pi^I\Box_y
+\f{1}{10}h^I\Box_y
+\f{\ti{\ap}}{2}b_I\Box_y-\f{\ti{\ap}^2}{5}\pi^I\right]Y^I=0.\label{EinABfour}
\ea

\subsubsection{$\mu\ap$ Components}

In a similar way, from $(\mu, \alpha)$-component of \eqref{ein}, we
obtain the following equations: \ba && \left[\f{1}{2}(\Box_x
B^I_\mu-\nabla^\nu\nabla_\mu B^I_\nu)+\f{1}{2}B^I_\mu\Delta_y
+\f{\ti{\ap}}{4}b^I_\mu\Delta_y+\f{\ti{\ap}}{4!}\ep_\mu^{\
\rho_1\rho_2\rho_3\rho_4}\nabla_{\rho_1}
b^I_{\rho_2\rho_3\rho_4}\right]Y^I_\ap=0.\label{EinMAone}\\
&& \Biggl[-\f{1}{2}\nabla^\nu h^I_{\nu\mu}+\f{1}{2}\nabla_\mu
h^I+\f{2}{5}\nabla_\mu\pi^I +\f{\ti{\ap}}{4}\nabla_\mu
b^I+\f{\ti{\ap}}{96}\ep_\mu^{\
\rho_1\rho_2\rho_3\rho_4}b^I_{\rho_1\rho_2\rho_3\rho_4}\no && \ \ \
\ \ \ \ \ \ \ \ +\delta_{\mu,w}\f{1}{2}e^{2\phi}\de_w \chi~ \eta^I
 +\delta_{\mu,r}\f{1}{2}\de_r\phi~ \vp^I\Biggr]\nabla_\ap Y^I=0.\label{EinMAtwo}
\ea

\subsubsection{$\mu\nu$ Components}

Finally, from $(\mu,\nu)$-component of \eqref{ein}, we obtain the
following equations: \ba
0&=&-\f{1}{2}(\Box_x+\Box_y)h_{\mu\nu}-\f{1}{2}\nabla_\mu\nabla_\nu
(h+\pi) +\f{1}{2}(\nabla_\mu\nabla^\rho
h_{\rho\nu}+\nabla_\nu\nabla^\rho h_{\rho\mu})
+R_{\mu\rho\sigma\nu}h^{\rho\sigma}\no &&+\f{1}{2}(R_\mu^\rho
h_{\nu\rho}+R_\nu^\rho h_{\mu\rho})
-e^{2\phi}(\de_\mu\chi\de_\nu\chi)\vp-\f{e^{2\phi}}{2}
(\de_\mu\chi\de_\nu\eta+\de_\nu\chi\de_\mu\eta)\no
&&-\f{1}{2}\de_\mu\phi\de_\nu \vp-\f{1}{2}\de_\nu\phi\de_\mu \vp
-\f{\ti{\ap}}{48}g_{\mu\nu}\ep^{\rho_1\rho_2\rho_3\rho_4\rho_5}
\nabla_{\rho_1}b^I_{\rho_2\rho_3\rho_4\rho_5}Y^I
-\f{\ti{\ap}^2}{4}g_{\mu\nu}h+\f{\ti{\ap}^2}{4}h_{\mu\nu}.\nonumber\\
\label{EinMM} \ea

\subsection{Dilaton and Axion Equations}
\label{subsec:linear_dilaton}

Let us then move to the linear perturbation for the dilaton and
axion equations of motion. They are given by \ba
&&\de_M(\s{-g}g^{MN}\de_N\phi)-\s{-g}e^{2\phi}g^{MN}\de_M\chi
\de_N\chi=0,
\\
&&\de_M(\s{-g}e^{2\phi}g^{MN}\de_N\chi)=0. \ea respectively. When
fluctuating around our background, these equations are rewritten as
\ba
&&(\Box_x+\Box_y)\vp+\f{1}{2}g^{rr}\de_r\phi\de_r(h+\pi)-(\de_r\phi)(\nabla_\mu
h^{\mu r})-(\nabla_I\de_J\phi)h^{IJ} \no &&\qquad\qquad\quad-2\vp
e^{2\phi} g^{ww}(\de_w\chi)^2 +e^{2\phi}h^{ww}(\de_w\chi)^2
-2e^{2\phi}g^{ww}(\de_w\chi)(\de_w\eta)=0, \label{dilaton_expanded}\\
&&(\Box_x+\Box_y)\eta+2(\de_r\phi)(\de_r\eta)g^{rr}+2g^{ww}(\de_w\vp)(\de_w\chi)
+\f{1}{2}g^{ww}\de_w(h+\pi)\de_w\chi \no
&&\quad\quad\quad\quad\quad\qquad\quad -(\nabla_\mu h^{\mu
w})(\de_w\chi)
-h^{IJ}(\nabla_I\de_J\chi)-2h^{rw}(\de_r\phi)(\de_w\chi)=0.
\label{axion_expanded} \ea

\subsection{Spectrum for Decoupled Modes}

As we derived the linearized equations of motion for the
fluctuations in the appendix \ref{subsec:linear_five},
\ref{subsec:linear_einstein}, and \ref{subsec:linear_dilaton}, we
then analyze the spectrum for them. We start with the one for those
modes which do not mix with other modes in a complicated way. The
analysis turns out to be essentially the same as the case of
$AdS_5\times S^5$ \cite{KRN}.

\subsubsection{$\phi^I$ Mode from $h_{(\ap\beta)}$}

First we consider the scalar mode
$h_{(\ap\beta)}=\phi^IY^I_{(\ap\beta)}$ from the fluctuation of the
metric. It obeys the equation of motion (\ref{EinABone}) and, by
using \eqref{harm_sym}, we obtain \be
\left(\Box_x-\f{k(k+4)}{R^2}\right)\phi^I=0.\ \ \ (k=2,3,4,\ddd) \ee
Thus we find its mass $m^2=\f{k(k+4)}{R^2}$, which obviously
satisfies the stability condition.

\subsubsection{$b^I_{\mu\nu}$ Mode from $C_{\mu\nu\ap\beta}$}

For $b^I_{\mu\nu}$ from the fluctuation of $C_{\mu\nu\ap\beta}$,
(\ref{ffluxthree}) leads to \be
(\mbox{Max}_x+\Delta_y)b^I_{\mu\nu}Y^I_{[\ap\beta]}=0. \ee The
Maxwell operator $\mbox{Max}_x$ on $M_5$ is defined by $\mbox{Max}_x
b^I_{\mu}= \Box_x b_{\mu_1} -\nabla^{\nu}\nabla_{\mu}b_{\nu}$ for
the vector $b^I_{\mu}$ and we can generalize the definition for
teonsors. Since the mass for this mode is given by
$m^2=-\Delta_y=\f{(k+2)^2}{R^2}\ $ $(k=1,2,\ddd)$, this mode turns
out to be stable.

\subsubsection{$b^I_{\mu}$ and $B^I_\mu$ Modes from $g_{\mu\ap}$ and $C_{\mu\ap\beta\gamma}$}

Let us next consider $b^I_{\mu}$ and $B^I_\mu$ from the fluctuation
of $g_{\mu\ap}$ and $C_{\mu\ap\beta\gamma}$ respectively. From
(\ref{ffluxtwo}), (\ref{EinMAone}), and (\ref{fluxfive}), we obtain
the equations for these modes as \ba &&((\mbox{Max}_x+\Delta_y)
b^I_\mu-\ti{\ap}B^I_\mu)Y^I_\ap=0,
\\
&&((\mbox{Max}_x+\Delta_y) B^I_\mu
-\f{\ti{\ap}}{2}(Max-\Delta_y)b^I_\mu)Y^I_\ap=0. \ea
or, by denoting in a different expression, as \ba \mbox{Max}_x\cdot
\left(
  \begin{array}{c}
    b^I_\mu \\
    B^I_\mu \\
  \end{array}
\right) + \left(
  \begin{array}{cc}
    \Delta_y & -\ti{\ap} \\
    \ti{\ap}\Delta_y &  \Delta_y-\f{\ti{\ap}^2}{2}\\
  \end{array}
\right)\cdot \left(
  \begin{array}{c}
    b^I_\mu \\
    B^I_\mu \\
  \end{array}
\right)=0. \ea The eigenvalues of $2\times 2$ matrix in the second
term are \ba
-m^2_\pm=\Delta_y-\f{\ti{\ap}^2}{4}\pm\s{\f{\ti{\ap}^4}{16}-\ti{\ap}^2\Delta_y}\
. \ea More explicitly, by using \eqref{harm_vector}, we get the
masses for these vector fields \ba m^2_+=\f{k^2-1}{R^2},\ \ \ \ \
m^2_-=\f{(k+3)(k+5)}{R^2},\ \ \ \ \  (k=1,2,3,\ddd). \ea Therefore
we can find that these modes are stable, too.

\subsection{Spectrum for Mixed Modes: $h^I$, $\pi^I$, and $b^I$}

Let us determine the spectrum for the scalar perturbation $h^I$,
$\pi^I$, and $b^I$ here. We first assume $k\geq 2$ and then find
$h^I=-\f{3}{5}\pi^I$ from (\ref{EinABthree}). Thus we can rewrite
(\ref{ffluxone}) and (\ref{EinABfour}) as follows: \ba \Box_x\cdot
\left(
  \begin{array}{c}
    b^I \\
    \pi^I \\
  \end{array}
\right) + \left(
  \begin{array}{cc}
    \Box_y & -\f{4}{5}\ti{\ap} \\
    5\ti{\ap}\Box_y &  \Box_y-2\ti{\ap}^2\\
  \end{array}
\right)\cdot \left(
  \begin{array}{c}
    b^I \\
    \pi^I \\
  \end{array}
\right)=0. \ea The matrix in the second term is diagonalized and the
eigenvalues are given by \be -m^2_{\pm}=\Box_y-\ti{\ap}^2\pm
\s{\ti{\ap}^4-4\ti{\ap}^2\Box_y}, \ee or, more explicitly, by \ba
&& m^2_+=\f{k^2-4k}{R^2},\ \ \ \   \label{plusmass}\\
&& m^2_-=\f{(k+4)(k+8)}{R^2}.\ \ \ \ .\label{mixmass} \ea Even for
$k=0,1$, we can see that the expression \eqref{mixmass} is correct.
Thus we find the lowest mass in this mode is $m_+^2=-\f{4}{R^2}$
when $k=2$. This violates the stability bound.

\subsection{Spectrum for Other Modes: $h_{\mu\nu}$, $\vp$, and $\eta$}

Now, the remaining modes are $h_{\mu\nu}$, $\vp$, and $\eta$.
Since the analysis of massive modes looks highly complicated, below
we would like to consider only zero modes on $S^5$ of $h_{\mu\nu}$
and $\vp,\eta$. They are useful to the calculations of the
viscosity in section \ref{sec:viscosity}. For this reason, we
generalize the background to the black brane metric \eqref{bhsol}
and write down the linearized equations for the fluctuation around
it. By taking $\mu\to 0$, we can reproduce those for the background
metric \eqref{scalingmetric}.

\subsubsection{Zero Modes on $S^5$}
\label{zeromodeh}

We concentrate on the zero modes on $S^5$ i.e. $k=0$ modes of the
spherical harmonics. Notice that in this case, we have $b=0$ and
$b_{\mu_1\mu_2\mu_3\mu_4}$ is expressed in terms of $h$ and $\pi$ as
\be
5\nabla_{\mu_1}b_{\mu_2\mu_3\mu_4\mu_5}=\f{2}{R}\ep_{\mu_1\mu_2\mu_3\mu_4\mu_5}
\left(h-\pi\right),\label{dualc} \ee from (\ref{selfdual}). From
(\ref{EinABfour}), the mode $\pi$ satisfies \be
\left(\Box_x-\f{32}{R^2}\right)\pi=0. \label{pic} \ee

It is also useful to define the Weyl shifted metric $H_{\mu\nu}$ by
\be H_{\mu\nu}=h_{\mu\nu}+\f{1}{3}g_{\mu\nu}\pi, \ee which
corresponds to the metric perturbation around the five-dimensional
background obtained by the reduction of $S^5$. Then the Einstein
equation (\ref{EinMM}) can be rewritten as follows: \be \delta
R^{(5)}_{\mu\nu}+\f{4}{R^2}H_{\mu\nu}-e^{2\phi}(\de_\mu\chi\de_\nu\chi)\vp-\f{e^{2\phi}}{2}
(\de_\mu\chi\de_\nu\eta+\de_\nu\chi\de_\mu\eta)
-\f{1}{2}(\de_\mu\phi\de_\nu \vp+\de_\nu\phi\de_\mu
\vp)=0.\label{einc} \ee Here $\delta R^{(5)}_{\mu\nu}$ is the
perturbation of the purely five dimensional Ricci tensor (neglecting
the $S^5$ contributions) due to the metric perturbation
$H_{\mu\nu}$. Notice that when $\mu,\nu\neq w,r$, we obtain the
simple Einstein equation $\delta
R^{(5)}_{\mu\nu}+\f{4}{R^2}H_{\mu\nu}=0$. This simplification is
applicable, for example, to the shear modes $H_{tx}$ and $H_{xy}$.

For the dilaton equation of motion \eqref{dilaton_expanded} with
$k=0$, by using the results \ba &&\nabla_t \de_t \phi=-\f{2}{3}r^2
\left(1-\f{\mu}{r^{11/3}}\right)\left(1+\f{5\mu}{6r^{11/3}}\right)\equiv
f_t,\no &&\nabla_x \de_x \phi=\nabla_y \de_y \phi=
\f{2}{3}r^2\left(1-\frac{\mu}{r^{11/3}}\right)\equiv f_x,\no
&&\nabla_w \de_w \phi=\f{4}{9}r^{\f{4}{3}}
\left(1-\frac{\mu}{r^{11/3}}\right)\equiv f_w,\no &&\nabla_r \de_r
\phi=\f{11\mu}{9r^{17/3}}\left(1-\frac{\mu}{r^{11/3}}\right)^{-1}\equiv
f_w, \ea and the Weyl shifted metric $H_{\mu\nu}$, we can rewrite it
as follows: \ba && \Box_x\vp+\f{1}{3r}g^{rr}(\de_r H)-f_t
H^{tt}-f_x(H^{xx}+H^{yy})-f_w H^{ww} -f_r H^{rr}\no
&&\qquad\qquad\qquad+\f{1}{3}\pi(g^{tt}f_t+2g^{xx}f_x+g^{ww}f_w+g^{rr}f_r)-\f{2}{3r}\nabla_\mu
H^{\mu r} \no &&\qquad\qquad\qquad +\beta^2
e^{2\phi}(H^{ww}-2g^{ww}\vp-\f{1}{3}\pi g^{ww}) -2\beta
e^{2\phi}g^{ww}\de_w \eta=0,\label{dilatonE} \ea where $H\equiv
H^\mu_\mu=h+\f{5}{3}\pi$.

On the other hand, for the axion equation of motion
\eqref{axion_expanded} with $k=0$,
by using the values \be
\nabla_r\de_w\chi=\nabla_w\de_r\chi=-\f{2k}{3Lr}, \ee it can be
rewritten as follows \ba \Box_x \eta +\f{4}{3r}g^{rr}\de_r\eta+\beta
g^{ww}\de_w \left(2\vp+\f{H}{2}\right)-\beta \nabla_\mu H^{\mu w}=0.
\label{axionE} \ea

We can take the radial gauge $H_{\mu r}=0$ and have 5 physical modes
for $H_{\mu\nu}$. Then the dilaton and axion equations of motion
(\ref{dilatonE}) and (\ref{axionE}) become \ba &&
\Box_x\vp+\f{1}{3r}g^{rr}(\de_r H)-f_t H^{tt}-f_x(H^{xx}+H^{yy}) \no
&&\quad\quad\quad-f_w H^{ww}
+\f{1}{3}\pi(g^{tt}f_t+2g^{xx}f_x+g^{ww}f_w+g^{rr}f_r)  \no
&&\qquad\qquad\qquad +\beta^2
e^{2\phi}(H^{ww}-2g^{ww}\vp-\f{1}{3}\pi g^{ww}) -2\beta
e^{2\phi}g^{ww}\de_w \eta=0,\label{dilatonEE} \ea and \ba \Box_x
\eta +\f{4}{3r}g^{rr}\de_r\eta+\beta g^{ww}\de_w\left(2\vp+\f{H}{2}
\right)-\beta \nabla_\mu H^{\mu w}=0. \label{axionEE} \ea

In summary, we need to solve the zero mode equations of motion
(\ref{dualc}), (\ref{pic}), (\ref{einc}), (\ref{dilatonEE}), and
(\ref{axionEE}) to find the variables $b_{\mu\nu\rho\sigma}$, $\pi$,
$\vp$, $\eta$, and $H_{\mu\nu}$.

\section{Linearized Equations for the Shear and Sound Channel}
\label{sec:viscosity_equation}

In this appendix we summarize the linearized equations for the shear
and sound channel. They are useful to derive the differential
equations for the gauge invariant combinations in section
\ref{sec:viscosity}.

For this purpose, we consider the equations
(\ref{dualc}),(\ref{pic}),(\ref{einc}),(\ref{dilatonEE}), and
(\ref{axionEE}) derived in the appendix \ref{zeromodeh}. Then we
substitute the fluctuation corresponding to the shear channel
\eqref{shear_ch} or sound channel \eqref{sound_ch} into them and
then derive explicit expressions for the linearized equations of
motion. We summarize the resulting equations below.

\subsection{Shear Channel}
\label{subsec:shear_eq}

By considering the fluctuations $H_{tx}$ and $H_{xy}$ which
correspond to the shear channel in the radial gauge $H_{\mu r}=0$,
we obtain the equations for them
\begin{eqnarray}
&&H''_{tx}+\ln'(c_x^{14/3})H'_{tx} -\frac{q}{c_x^2c_t^2}
(qH_{tx}+\omega H_{xy})=0,  \label{shear1}\\
&&H_{xy}''+\ln'(c_t^2c_x^{8/3})H'_{xy} +\frac{\omega}{c_t^4}
(qH_{tx}+\omega H_{xy}) = 0,  \label{shear2}\\
&&qH_{xy}'+\omega \frac{c_x^2}{c_t^2}H_{tx}'=0, \label{shear3}
\end{eqnarray}
from $(t,x)$-, $(x,y)$-, and $(x,r)$-component of \eqref{einc},
respectively. We can directly show that two of the three equations
are independent.

\subsection{Sound Channel}
\label{subsec:sound_eq}

Let us next consider the sound mode. The corresponding fluctuations
are $H_{tt}, H_{ty}, H_{xx},H_{yy},H_{ww},\varphi,\pi$ and
$f_{\mu_1\mu_2\mu_3\mu_4\mu_5}$ in the radial gauge. In this case
the linearized equations (\ref{pic}), (\ref{einc}), and
(\ref{dilatonEE}) are summarized as
\begin{eqnarray}
&&H_{tt}''+\ln'\left(c_{t}^3c_{x}^{8/3}\right)H_{tt}'-\ln'\left(c_t\right)
(H_{yy}'+H_{ii}')  \nonumber \\
&&\qquad\qquad\qquad\qquad\qquad\quad-\frac{1}{c_{t}^2}\left(\frac{\omega^2}{c_{t}^2}(H_{yy}+H_{ii})
+\frac{2q\omega}{c_{t}^2}H_{ty}+\frac{q^2}{c_{x}^2}H_{tt}\right)= 0, \\
&&H_{xx}''+\ln'\left(c_{t}^2c_{x}^{8/3}\right)H'_{xx}
-\ln'(c_{x})\left(H_{tt}'-H_{yy}'-H_{ii}'\right)
+\frac{1}{c_{t}^2}\left(\frac{\omega^2}{c_{t}^2}-\frac{q^2}{c_{x}^2}\right)H_{xx}=0, \\
&&H_{yy}''+\ln'(c_{t}^2c_{x}^{8/3})H_{yy}'-\ln'(c_x)(H_{tt}'-H_{yy}'-H_{ii}') \nonumber\\
&&\qquad\qquad\qquad\qquad\qquad\qquad +\frac{1}{c_{t}^2}
\left(\frac{\omega^2}{c_{t}^2} H_{yy}
+\frac{2q\omega}{c_{t}^2}H_{ty}+\frac{q^2}{c_{x}^2}(H_{tt}-H_{ii})
\right)=0,\\
&&H_{ww}''+\ln'(c_t^2c_x^{8/3})H_{ww}'-\ln'(c_x^{2/3})(H_{tt}'-H_{yy}'-H_{ii}') \nonumber\\
&&\qquad\qquad\qquad\qquad\qquad\qquad
+\frac{1}{c_{t}^2}\left(\frac{\omega^2}{c_{t}^2}
-\frac{q^2}{c_{x}^2}\right)H_{ww}+\frac{44}{9c_{t}^2}\left(\varphi-\frac{1}{2}H_{ww}\right)=0,\\
&&(H_{tt}''-H_{yy}''-H_{ii}'')+\ln'(c_t c_x^2)(H_{tt}'-H_{yy}'-H_{ii}') \nonumber\\
&&\qquad\qquad\qquad\qquad\qquad\qquad +\ln'(c_t^2c_x^{-2})
H_{tt}'-\ln'c_x^{4/3}
\left(\varphi'-\frac{1}{2}H_{ww}'\right)=0, \\
&&H_{ty}''+ \ln'(c_{x}^{14/3})H_{ty}'+\frac{q\omega}{c_{t}^2c_{x}^2}H_{ii}=0,\\
&&(H_{yy}'+H_{ii}')+\frac{q}{\omega}H_{tz}'
+\ln'(c_{x}c_{t}^{-1})(H_{yy}+H_{ii})\nonumber\\
&&\qquad\qquad\qquad\qquad\qquad\qquad+\frac{q}{\omega}\ln'(c_x^2c_t^{-2})H_{ty}+
\ln(c_{x}^{2/3})\left(\varphi-\frac{1}{2}H_{ww}\right) = 0,\\
&& (H_{tt}'-H_{ii}')+\frac{\omega}{q}\frac{c_x^2}{c_t^2}H_{tz}'
-\ln'(c_xc_t^{-1})H_{tt}-\ln'(c_x^{2/3})\left(\varphi-\frac{1}{2}H_{ww}\right)=0,\\
&& \varphi''+\ln'(c_t^2 c_x^{8/3})\varphi'
-\ln'(c_x^{1/3})(H_{tt}'-H_{yy}'-H_{ii}')
+\frac{1}{c_{t}^2}\left(\frac{\omega^2}{c_t^2}-\frac{q^2}{c_x^2}\right)
\varphi \nonumber\\
&&\quad\quad\quad\quad\quad\quad\qquad
-\frac{44}{9c_{t}^2}\left(\varphi-\frac{1}{2}H_{ww}\right)+
\frac{11(\mu^3-2\mu^2r^{11/3}+2r^{11})}{27r^{28/3}(r^{11/3}-\mu)}\pi=0,
\nonumber \\
&&\pi''+\ln'(c_t^2 c_x^{8/3})\pi'
+\frac{1}{c_{t}^2}\left(\frac{\omega^2}{c_t^2}-\frac{q^2}{c_x^2}\right)
\pi-\frac{32}{c_t^2}\pi=0,
\end{eqnarray}
where $H_{ii} = H_{xx} + H_{ww}$.

\section{D3-D5 Scaling Solution with F-string Sources}
\label{sec:baryon}

Here we briefly review the D3-D5 scaling solution in type IIB supergravity
with F-string sources \cite{ANT}.
This solution can be regarded as a back-reacted
solution dual to a homogeneous baryon condensation in ${\cal N}=4$ super Yang-Mills.
Each baryon (= D5-brane) situated near the horizon $r=0$ carries
$N$ F-strings which extend into the boundary $r=\infty$ due to the string creation \cite{Witten}.
Below, we work in the Einstein frame
and the supergravity action is obtained by rewriting the one
in the string frame \eqref{sugra_action} by the
Einstein frame metric $G_{MN}^{E}=e^{-\phi/2}G_{MN}^{string}$.

For this brane setup, it is appropriate to use the following ansatz
for the metric \be
ds^2=-e^{u(r)}A(r)dt^2+e^{b(r)}\left(\sum_{i=1}^3(dx_i)^2\right)
+e^{c(r)}(A^{-1}(r)dr^2+r^2ds_{X_5}^2), \label{ansazp} \ee
where $X_5$ represents a Einstein manifold with the same
Ricci curvature as the unit radius $S^5$. For fluxes, the ansatz is
\ba && F_{3}=e^{\f{3}{2}b(r)}\cdot h(r)\cdot dx_1\wedge dx_2\wedge
dx_3, \no &&
F_5=r^5e^{\f{5}{2}c(r)}f(r)(\Omega_{X_5}+*\Omega_{X_5}),\ea 
where $\Omega_{X_5}$ is the volume form on $X_5$. 
We also assume that the dilaton is
dependent on the radial coordinate $r$ only. If we treat this
solution within supergravity, we will have a tadpole of the NSNS
3-form flux $H_3$ as is clear from the equation of motion in the
presence of the Chern-Simons term \be d(e^{-\phi}*H_3)=F_3\we
F_5,\label{CSeq} \ee where $F_3$ and $F_5$ are sourced by the D5 and
D3-branes, respectively. Thus we cannot construct consistent
supergravity solutions under the assumption of symmetry of spacetime
implied by (\ref{ansazp}).

To resolve this problem, we notice
that the effective F-string charges are generated from the
D3-brane and D5-brane charges via the Chern-Simons term
\begin{eqnarray}
S_{CS} =-\frac{1}{4\kappa_{10}^2}\int C_4\wedge H_3\wedge F_3.
\end{eqnarray}
Therefore, to resolve this problem, we treat the created F-strings
as the probe action and add it to the supergravity action. The probe
action for a single F-string in the Einstein frame is given by \be
S_{string}= -\f{1}{2\pi}\int d\tau d\sigma
e^{\f{\phi}{2}}\s{-G_{E}}+\f{1}{2\pi}\int d\tau d\sigma
B_{\mu\nu}\de_{\tau}X^\mu \de_{\sigma}X^\nu. \ee By identifying
$\tau=t$ and $\sigma=r$ and taking the sum over infinitely many
F-strings, we obtain the probe action for the created F-strings
 \be \sum_{i}S_{string(i)}=\int dx_1dx_2dx_3 \Omega_{X_5}\ \rho\
S_{string}, \ee where $i$ labels $i$-th F-string and the density of
F-strings $\rho$ is assumed to be constant.
In the following, we solve the equations of motion derived from
the supergravity action with this probe action added.

\subsection{Equation of Motions}

The Bianchi identity for $F_3$ and the equation of motion for
$F_5$ are written as
\be
\de_{r}(r^5e^{\f{5}{2}c(r)}f(r))=0, \ \ \ \ \
\de_r(e^{\f{3}{2}b(r)}h(r))=0. \ee
Here we notice that the equation of motion for $F_3$ is automatically
satisfied for the ansatz introduced above.
From these equations, we can define the constants $F$ and $H$ as follows: \be
f(r)=F\cdot r^{-5} e^{-\f{5}{2}c(r)},\ \ \ \ \ \ \ h(r)=H\cdot
e^{-\f{3}{2}b(r)}. \ee
Then, from (\ref{CSeq}), the density of F-strings turns out to be \be \rho=\f{\pi
FH}{2\kappa^2_{10}}. \ee

To derive the dilaton equation of motion, we notice that the F-string
action now looks like \be S_{F-string}=-\f{1}{2\pi}\int
dtdre^{\f{\phi}{2}}\s{-G_E}=-\f{1}{2\pi}\int
dtdre^{\f{\phi}{2}+\f{c}{2}+\f{u}{2}}. \ee Thus the equation of
motion becomes \be
\de_r(\phi'(r)A(r)e^{2c(r)+\f{3}{2}b(r)+\f{u(r)}{2}}r^5)=\f{FH}{4}
e^{\f{\phi(r)}{2}+\f{c(r)}{2}+\f{u(r)}{2}}+\f{1}{2}
r^5e^{\phi(r)+\f{3}{2}b(r)+3c(r)+\f{u(r)}{2}} h(r)^2.
\label{bdilaton}\ee

We can also derive the Einstein equations for the type IIB
supergravity with the F-string action. Combined with
(\ref{bdilaton}), we can summarize the equations of motion as
follows: \ba && [A(\log
A+u)'r^5e^{\f{3}{2}b+2c+\f{1}{2}u}]'=\f{3}{8}FHe^{\f{1}{2}\phi+\f{1}{2}c+\f{1}{2}u}
+\f{1}{2}F^2r^{-5}e^{-2c+\f{3}{2}b+\f{1}{2}u}
+\f{1}{4}H^2r^5e^{\phi+3c-\f{3}{2}b+\f{1}{2}u},\no  \label{eoma} \\
&& [b'A
r^5e^{\f{3}{2}b+2c+\f{1}{2}u}]'=-\f{1}{8}FHe^{\f{1}{2}\phi+\f{1}{2}c+\f{1}{2}u}
+\f{1}{2}F^2r^{-5}e^{-2c+\f{3}{2}b+\f{1}{2}u}
-\f{3}{4}H^2r^5e^{\phi+3c-\f{3}{2}b+\f{1}{2}u}, \label{eomb} \\ &&
[(c+2\log
r)'Ar^5e^{\f{3}{2}b+2c+\f{1}{2}u}]'=-\f{1}{8}FHe^{\f{1}{2}\phi+\f{1}{2}c+\f{1}{2}u}
-\f{1}{2}F^2r^{-5}e^{-2c+\f{3}{2}b+\f{1}{2}u}\no &&\ \ \ \ \ \ \ \ \
\ \ \ \ \ \ \ \ \ \ \ \ \ \ \ \ \ \ \ \ \ \ \ \ \ \ \ \ \ \ \ \ \ \
+\f{1}{4}H^2r^5e^{\phi+3c-\f{3}{2}b+\f{1}{2}u}
+8r^3e^{2c+\f{3}{2}b+\f{1}{2}u},\label{eomc} \\ && [(\log
A+u-b-2\phi)'Ar^5e^{\f{3}{2}b+2c+\f{1}{2}u}]'=0,\label{eomd} \\ &&
2(\phi')^2+3(b')^2
+10\f{c'}{r}-3b'c'+6b''+10c''-3b'u'-5c'u'-10\f{u'}{r}=0.
\label{eome} \ea

It is also useful to derive the following equation from a linear
combination of (\ref{eoma}), (\ref{eomb}), and (\ref{eomc}) \be
[(\log A+u+b+2c+4\log r)'Ar^5 e^{\f{3}{2}b+2c+\f{1}{2}u}]'=16r^3
e^{2c+\f{3}{2}b+\f{1}{2}u}. \label{eomf} \ee

In the above discussion, we have derived five equation of motion for
five variables $A(r)$, $u(r)$, $b(r)$, $c(r)$, $\phi(r)$. However,
we can eliminate one of them, say $u(r)$, by the diffeomorphism
$r\to \ti{r}=\ti{r}(r)$. In this sense, the independent degree of
freedom under the symmetry ansatz is the four variables. Thus we
should show the five equations of motion are not over constrained.
Indeed we can show the following identity from the four equations of
motion (\ref{eoma}), (\ref{eomb}), (\ref{eomc}), and (\ref{eomd})
\be \left[r^{10}A(r)^2e^{3b(r)+4c(r)+u(r)}\cdot
E_{constraint}(r)\right]'=0, \ee where $E_{constraint}(r)$ is the
left-hand side of (\ref{eome}). This guarantees that the fifth
equation of motion $E_{constraint}(r)=0$ is satisfied if it is
vanishing at any particular value of $r$.

\subsection{Scaling Solutions}
Let us assume the following simple scaling profile for the
unknown functions:  \be \  u(r)=u_1\log r+u_0,\ \ \ b(r)=b_1\log
r+b_0,\ \ c(r)=c_1\log r +c_0, \ \ \phi(r)=\phi_1\log r +\phi_0.
\label{anslog} \ee At first we also assume $A(r)=1$, which
corresponds to the extremal case.

Using the three equations of motion (\ref{eomc}), (\ref{eomd}), and
(\ref{eome}), we find that the following coefficients $u_1,b_1,c_1$
and $\phi_1$ satisfy equations of motion \be u_1=7\s{\f{2}{5}},\ \
b_1=\s{\f{2}{5}},\ \ c_1=-2,\ \  \phi_1=3\s{\f{2}{5}}.
\label{values} \ee By substituting \eqref{anslog} with
\eqref{values} into (\ref{eoma}), (\ref{eomb}) and (\ref{eomc}), we
can find four solutions for the pair ($F$, $H$) in terms of
$u_0,b_0,c_0$ and $\phi_0$ as follows: \ba F&=&\pm e^{2c_0}\cdot
\f{\s{83-\s{1081}}}{\s{3}}, \no H&=& ^{}\f{\s{1081}-5}{44}\cdot
e^{\f{3}{2}b_0-\f{5}{2}c_0-\f{1}{2}\phi_0}\cdot F, \label{fluxr1}
\ea and \ba F&=&\pm e^{2c_0}\cdot \f{\s{83+\s{1081}}}{\s{3}}, \no
H&=&- ^{}\f{\s{1081}+5}{44}\cdot
e^{\f{3}{2}b_0-\f{5}{2}c_0-\f{1}{2}\phi_0}\cdot F. \label{fluxr2}
\ea Thus, in order to realize the fluxes with arbitrary values and
sign, we have only to choose one of the four solutions and tune
$c_0$ and $\phi_0$ appropriately. We set $u_0=b_0=0$ below.

As we constructed the extremal solution, generalization to a black
brane solution is straightforward by considering more general
functions $A(r)$. Let us again assume the ansatz (\ref{anslog}) with
the values (\ref{values}). Then it immediately follows that all
equation of motion are satisfied only if \be
A(r)=1-\f{M}{r^{\sqrt{10}}},\ \ \ \ee where $M$ is an arbitrary
constant related to the ADM mass of the black brane. We also notice
that the other profiles are the same as the extremal solutions.

In summary we obtain the following metric in the Einstein frame and
the dilaton
 \be
ds^2=-r^{u_1}A(r)dt^2+r^{b_1}d\vec{x}^2+e^{c_0}\left(\f{dr^2}{r^2A(r)}
+(d\Omega_5)^2\right), \ \ \ \ e^{\phi(r)-\phi_0}=r^{\phi_1}.
\label{solon} \ee
and, after the redefinition of radial coordinate $\rho=r^{\f{1}{\s{10}}}$,
we reach the expression (\ref{baryonm}).



\end{document}